\documentstyle[preprint,aps]{revtex}
\tighten
\draft
\begin{document}
\preprint{\today}
\title{Dynamics of the rotational degrees of freedom in a supercooled
liquid of diatomic molecules}
\author{Stefan K\"ammerer, Walter Kob\cite{wkob} and Rolf 
Schilling\cite{rschilling}}
\address{Institut f\"ur Physik, Johannes Gutenberg-Universit\"at,
Staudinger Weg 7, D-55099 Mainz, Germany}
\maketitle

\begin{abstract}
Using molecular dynamics computer simulations, we investigate the
dynamics of the rotational degrees of freedom in a supercooled system
composed of rigid, diatomic molecules. The interaction between the
molecules is given by the sum of interaction-site potentials of the
Lennard-Jones type.  In agreement with mode-coupling theory (MCT), we
find that the relaxation times of the orientational time correlation
functions $C_1^{(s)}(t)$, $C_2^{(s)}(t)$ and $C_{1}(t)$ show at low
temperatures a power-law with the same critical temperature $T_c$, and
which is also identical to the critical temperature for the
translational degrees of freedom.  In contrast to MCT we find, however,
that for these correlators the time-temperature superposition principle
does not hold well and that also the critical exponent $\gamma$ depends
on the correlator.  We also study the temperature dependence of the
rotational diffusion constant $D_r$ and demonstrate that at high
temperatures $D_r$ is proportional to the translational diffusion
constant $D$ and that when the system starts to become supercooled the
former shows an Arrhenius behavior whereas the latter exhibits a
power-law dependence. We discuss the origin for the difference in the
temperature dependence of $D$ (or the relaxation times of
$C_{l}^{(s)}$) and $D_r$. Finally we present results which show that at
low temperatures $180^{\circ}$ flips of the molecule are an important
component of the relaxation dynamics for the orientational degrees of
freedom.

\end{abstract}

\narrowtext


\section{Introduction}
\label{sec1}

In the last ten years mode coupling theory
(MCT)~\cite{gotze91,gotze92,schilling94,kob97} has led to a strong
interest in the phenomenon of the glass transition, since this theory
makes detailed predictions on the dynamics of glass formers in the
vicinity of the glass transition and thus challenges the
experimentalists to test these predictions. Starting from the
microscopic equations of motion, MCT derives, by using certain
approximations which are believed to be quite reliable for {\it simple}
liquids, an equation of motion for the density correlator. One of the
main results of MCT is the existence of a dynamical transition at a
temperature $T_c$, at which, in the so-called idealized version of the
theory, the dynamics of the system undergoes a transition from an
ergodic ($T > T_c$) to a nonergodic behavior ($T < T_c$), and which can
be interpreted as a glass transition.  Due to the presence of
ergodicity-restoring processes, commonly called ``hopping processes'',
most glass formers show to a certain extent deviations from the
predictions of the idealized theory, since this version of MCT does not
take into account these sort of processes. In the extended version of
the theory hopping processes are taken care of and one finds that in
the vicinity of $T_c$ a crossover in the behavior of the dynamics can
still be observed~\cite{gotze91,gotze92,kob97}.  Many of the predictions of
MCT were confirmed by experiments and in numerical simulations on
various glass forming systems. For reviews the reader is referred to
Refs.~\cite{gotze91,gotze92,schilling94,kob97,yip95}.

In the real world most of the good glass formers are molecular
systems.  One of the important differences between molecular systems
and simple liquids is that the former have orientational degrees of
freedom (ODOF). The dynamics of these ODOF can be measured, e.g., by
dielectric spectroscopy, light scattering, and NMR. For the molecular
glass formers salol, e.g., it has been shown that MCT gives a
satisfactory description of the relaxation behavior of these
systems~\cite{li92}. This conclusion seems to contradict the outcome of
experiments in which the dielectric response of this system was
probed~\cite{dixon90} because of the following fact: MCT
predicts that for all observables that couple to the density
fluctuations the imaginary part of the corresponding susceptibility
exhibits a minimum at the same frequency $\omega_{min}$, if one is very
close to $T_c$.  Light scattering experiments have shown that in salol
this minimum occurs at a frequency of about 2GHz if $T=253$
K~\cite{li92}. However, Dixon {\it et al.} showed that no such minimum
is observed in the dielectric function $\epsilon '' (\omega)$ for
frequencies up to 18GHz for $T=255$ K~\cite{dixon90} thus seemingly
contradicting the conclusions of Ref.~\cite{li92} that MCT gives a
correct description of the dynamics of this system.  Similar
conclusions were drawn for glycerol. However, in recent extensive
dielectric measurements on glycerol this minimum was found and it was
shown that the data can be described by the $\beta$-correlator of MCT
for simple liquids~\cite{lunkenheimer96,lunkenheimer962}, although in
this experiment the location of the minimum is indeed different from
the one in the light scattering experiment. This could be due to at
least two reasons: First, the MCT-equations for {\it molecular} liquids
might be different from those for {\it simple} liquids, hence leading
to different predictions. Second, the predictions of MCT for simple
liquids can be used, but corrections to the asymptotic laws of MCT have
to be taken into account. These corrections make the reduction theorem
invalid and therefore the location of the minimum depends on the
observable~\cite{franosch97a}.  We also note that very recently
evidence has been given that a minimum exists in $\epsilon''(\omega)$
for salol also~~\cite{lunkenheimer963}. However, its location remains
still undetermined

The above discussion shows that the role of the ODOF for the glassy
dynamics in the supercooled regime is certainly not settled and thus
remains an interesting field of research. One possibility to gain some
insight into the dynamics of supercooled liquids is to perform
molecular dynamics computer simulations of such
systems~\cite{kob_stauf}.  Simulations of {\it simple} liquids have
shown that this method can indeed be very useful to understand the
dynamics down to temperatures close to
$T_c$~\cite{roux_wahnstrom,kob94,kob95b}.  For {\it supercooled}
molecular liquids, however, only few numerical simulation have been
done so far. Signorini {\it et al.} investigated the dynamics of the
ionic glass former $[Ca(NO_3)_2]_{0.4} [KNO_3]_{0.6}$ (CKN), where the
$NO_3$ unit was treated as a rigid molecule~\cite{signorini90} and
Sindzingre and Klein studied methanol~\cite{sindzingre92}. OTP was
investigated by Lewis and Wahnstr\"om~\cite{lewis94,wahnstrom97}, who
modelled the molecule with a rigid isoscale triangle of Lennard-Jones
particles, and by Kudchadkar and Wiest~\cite{kudchadkar95} who used a
18-site, three-ring model. Very recently Sciortino {\it et al.}
presented their results on supercooled H$_2$O (also taken as a rigid
molecule)~\cite{sciortino96,sciortino_mrs}. In these papers a two-step
relaxation process, as predicted by MCT, was observed for both, the
translational and rotational degrees of freedom.  But a more
quantitative analysis in the framework of MCT was essentially
restricted to the translational degrees of freedom (TDOF).  In
particular it does not become clear from these studies whether the ODOF
freeze at the same temperature as the TDOF do since no detailed
analysis with respect to this point was made (although very recently
evidence was given, that for water the freezing temperatures of the
ODOF and the TDOF are very close together~\cite{sciortino_mrs}).  The
investigation of this point is one of the major objectives of the
present work. For this we will focus on the time scale of the
$\alpha$-relaxation and discuss our findings regarding the
$\beta$-relaxation elsewhere~\cite{kammerer97}.  We also mention that
in a very recent paper Ma and Lai investigated the dynamics of the
translational degrees of freedom for a dumb-bell shaped molecule and
argued that the molecular character leads to a decrease of the MCT
parameter $\lambda$~\cite{ma97}.

The molecular system we consider consists of molecules in which two
different atoms are connected rigidly.  Apart from diatomic molecules
with head-tail symmetry, this is the {\it simplest} molecular system
one can choose.  Although the choice of a linear molecule is of course
somewhat special, it is nevertheless interesting to study the dynamics
of such a molecular system in the supercooled regime in order to check
whether even such a simple system shows the phenomenon of a glass
transition.  Furthermore the continuous rotational symmetry around the
long axis of the molecules simplifies also the theoretical description
of this system, as compared to the molecules studied in Refs.
\cite{signorini90,sindzingre92,lewis94,kudchadkar95,sciortino96,sciortino_mrs}
and thus allows to make a more stringent test of the theory.  Although
the MCT equations for a diatomic molecule in a simple
liquid~\cite{franosch97b} and for molecular liquids have recently been
derived~\cite{scheidsteger97,schmitz97}, they have not been analyzed in
great detail so far. Since the structure of these equations is not
identical to that for simple liquids, it is not obvious whether the
predictions derived by G\"otze~\cite{gotze85} for MCT equations which
apply to simple liquids are still valid. Thus one of the main
motivation of the present work is to test whether these predictions
hold also for the molecular system investigated here.

The outline of our paper is as follows. In the next section we define
the various correlators and diffusion constants that we will study. In
addition we also summarize those results of MCT whose validity we will
check for the present system.  In section~\ref{sec3}  the model as well
as the details of the computer simulation are presented. The results
are given in Sec.~\ref{sec4} and are discussed in the final section in
which we also offer our main conclusions.

\section{Correlation functions and predictions of MCT}
\label{sec2}

In this section we define the correlation functions that are studied
in this work and recapitulate some of the predictions of MCT.

To study the static and dynamical properties of macroscopic systems
with $N$ particles it is convenient to use correlation functions. For
the translational degrees of freedom it is customary to characterize
the dynamics with the help of the van Hove correlation function, or
its space Fourier transform, the intermediate scattering function.
For the rotational degrees of freedom convenient correlation
functions are the functions $C_l(t)$ and $C_l^{(s)}(t)$ which are
defined as follows:

\begin{equation}
C_l(t) = \frac{1}{N} \sum_{n,n'} \langle
           P_{l}(\vec{u}_n(t) \cdot \vec{u}_{n'}(0)) \rangle
\qquad l\geq 1\qquad .
\label{eq2}
\end{equation}
and its self part
\begin{equation}
C_l^{(s)}(t) = \frac{1}{N} \sum_{n=1}^N \langle
           P_{l}(\vec{u}_n(t) \cdot \vec{u}_{n}(0)) \rangle 
\qquad l\geq 1 \quad .
\label{eq1}
\end{equation}
Here $P_{l}(x)$ is the Legendre polynomial of order $l$ and $\vec{u}_n$
is the normalized orientational vector of the $n$-th molecule. These
orientational correlation functions are straightforward generalizations
of the intermediate scattering function to the case where also ODOF are
present. More details on this point are given in
Ref.~\cite{kammerer97}. The experimental relevance of the functions
$C_l(t)$ is given by the fact that for $l=1$ and $l=2$ they can
be measured in dielectric and light scattering experiments,
respectively. We also note that it is often assumed (cf., e.g.,
Ref.~\cite{cummins96}) that the cross terms in $C_l(t)$ can be
neglected. In that case the experiments would also yield information on
$C_l^{(s)}$. We will discuss this point below.

Besides the intermediate scattering function or the correlators given
in Eqs.~(\ref{eq2}) and~(\ref{eq1}), the dynamics of the system can
also be studied by means of the auto-correlation functions of the
velocities $\vec{v}_n(t)$ of the particles

\begin{equation}
\Phi(t) = \frac{1}{N} \sum_{n=1}^{N} \langle
\vec{v}_n(t) \cdot \vec{v}_n(0) \rangle
\label{eq3}
\end{equation}
or the corresponding one of the angular velocities $\vec{\omega}_n(t)$
\begin{equation}
\Psi_l(t) = \frac{1}{N} \sum_{n=1}^{N} \langle P_l(\cos \alpha_n(t))\rangle
\qquad  l \ge 1 \qquad ,
\label{eq4}
\end{equation}
where $\alpha_n(t)$ is the angle between $\vec{\omega}_n(t)$ and
$\vec{\omega}_n(0)$.  It was recently shown that $\Psi_2(t)$ is
particularly useful to study the freezing of the ODOF~\cite{renner95}.
Since the translational and rotational diffusion constants, $D$ and
$D_r$, respectively, can be obtained by means of a Green-Kubo relation
from Eqs.~(\ref{eq3}) and (\ref{eq4}), respectively, the significance
of these time correlation functions is obvious.

In the strongly supercooled regime $\Phi(t)$ and $\Psi_1(t)$ exhibit a
negative, slowly decaying long time tail.  This makes the numerical
calculation of $D$ and $D_r$ from the Green-Kubo relation a difficult
task, if the temperatures are low.  Therefore it is customary to use
the corresponding Einstein relations, which are mathematically
equivalent to the Green-Kubo identity. For the TDOF $D$ is then
determined from the mean squared displacement:

\begin{equation}
D = \lim_{t\to\infty} \frac{1}{6t N} \sum_{n=1}^{N} \langle 
|\vec{x}_n(t) - \vec{x}_n(0)|^2 \rangle \qquad.
\label{eq5}
\end{equation}
To obtain the analogon to this equation for the ODOF we replace
$\vec{x}_n(t)$ by the corresponding variable $\vec{\phi}_n(t)$, which
is defined as

\begin{equation}
\vec{\phi}_n(t) - \vec{\phi}_n(0) = \Delta \vec{\phi}_n(t)= \int_0^t dt'
\vec{\omega}_n(t') \quad.
\label{eq6}
\end{equation}

In analogy to Eq.~(\ref{eq5}) we thus obtain the following Einstein 
relation for the ODOF:

\begin{equation}
D_r = \lim_{t\to\infty} \frac{1}{4t N} \sum_{n=1}^{N} 
\langle |\vec{\phi}_n(t) - \vec{\phi}_n(0)|^2 \rangle \qquad .
\label{eq7}
\end{equation}

Note that $\vec{\phi}_n(t)$ is not bounded, in contrast to $\vec{u}_n(t)$,
which is restricted to the surface of a unit sphere. 
This is the reason, why a replacement of $\vec{x}_n(t)$  by 
$\vec{u}_n(t)$ in Eq.~(\ref{eq5}) would yield $D_r = 0$.

Let us now recapitulate those predictions of the idealized version of
MCT which are relevant for the present work. More detail can be found
in Refs.~\cite{gotze91,gotze92,schilling94,kob97}.  The theory predicts
that, in the vicinity of the critical temperature $T_c$, all time
correlation functions $\phi(t)$ which couple to the density correlation
function should show a two-step relaxation behavior, i.e. exhibit a
plateau-like region when plotted versus the logarithm of time. The time
window in which $\phi$ is in the vicinity of this plateau is called the
$\beta$-relaxation regime. The time window in which the correlator
falls below this plateau is called the $\alpha$-relaxation regime.

MCT predicts that upon approaching $T_c$ from above, the 
$\alpha$-relaxation time $\tau(T)$ diverges with a power-law, i.e.

\begin{equation}
\tau(T) \propto (T-T_c)^{-\gamma} \qquad,
\label{eq8}
\end{equation}
with a critical exponent $\gamma > 1.5$. Note that the values of 
$T_c$ and $\gamma$ are predicted to be independent of the correlator.
Furthermore the theory predicts that also the translational 
diffusion constant $D$ shows in the vicinity of $T_c$ a power-law
behavior, i.e.

\begin{equation}
D(T) \propto (T-T_c)^{\gamma} \qquad,
\label{eq9}
\end{equation}
with the same $\gamma$ as in Eq.~(\ref{eq8}). 

Finally the theory makes the prediction that the correlators should
obey the so-called time-temperature superposition principle (TTSP).
This means that if a correlator $\phi(t,T)$ is plotted versus the
reduced time $t/\tau(T)$, the curves corresponding to the different
temperatures fall, in the $\alpha$-relaxation regime, on a master
curve $\hat{\phi}$, i.e.

\begin{equation}
\phi(t,T)=\hat{\phi}(t/\tau(T))\qquad,
\label{eq10}
\end{equation}
the shape of which is approximated well by a
Kohlrausch-Williams-Watts function, i.e. $\hat{\phi}(t/\tau) \approx
A \exp( -(t/\tau)^{\beta})$, where the amplitude $A$ and the exponent
$\beta$ are {\it not} universal, i.e. will depend on $\phi$.

\section{Model and Details of the Simulation}
\label{sec3}

The model we investigate is a one-component system of rigid, diatomic
molecules. Each molecule is composed of two different Lennard-Jones
particles, in the following denoted by $A$ and $B$, which are separated
by a distance $d$ and each of which has the same mass $m$. The
interaction between two molecules is given by the sum of the
interaction between the four particles which is given by the
Lennard-Jones potential $V_{\alpha\beta}(r)=4\epsilon_{\alpha\beta}
\{(\sigma_{\alpha\beta}/r)^{12}-(\sigma_{\alpha\beta}/r)^{6}\}$ were
$\alpha,\beta \in \{A,B\}$. The Lennard-Jones parameters are given by
$\sigma_{AA}=\sigma_{AB}=1.0$, $\sigma_{BB}=0.95$,
$\epsilon_{AA}=\epsilon_{AB}=1.0$ and $\epsilon_{BB}=0.8$ and were
chosen such that the system did not show any sign of crystallization
even at the lowest temperatures investigated here. In the
following we will use reduced units and use $\sigma_{AA}$ as the unit
of length, $\epsilon_{AA}$ as the unit of energy (setting $k_B=1$) and
$(\sigma_{AA}^2m/48\epsilon_{AA})^{1/2}$ as the unit of time.  If the
atoms are argon-like this time unit corresponds to approximately 0.3ps.

The choice of the intramolecular distance $d$ between the $A$ and $B$
particles requires some consideration. On the one hand $d$ has to be
large enough to allow for a sufficiently strong coupling between the
translational and rotational degrees of freedom. On the other hand it
has to be so small that firstly the formation of liquid crystalline
structures is unlikely~\cite{allen87} and that secondly the energy
barrier involved in the intersection of two molecules is so large, that
at the temperatures and on the time scale of the simulation such a
crossing does not occur. We found that a value of 0.5 is a good
compromise.

In order to make the simulation more realistic we did it at constant
external pressure $p_{ext}$=1.0. For this we equilibrated the system in
the ($N,p,T$) ensemble, using the algorithm proposed by
Andersen~\cite{andersen81} and by setting the mass of the piston to
0.05.  The length of these equilibration runs always exceeded the
typical relaxation time of the system at the temperature considered,
which allows us to conclude that in the subsequent production runs we
were investigating the equilibrium dynamics of the system.  After
having determined from this equilibration run the appropriate density
of the system for the temperature of interest, we fixed the total
density to the so obtained density and started a production run in the
microcanonical ensemble using the rattle algorithm~\cite{andersen83}.
Note that it is advisable to make the production run in the
microcanonical ensemble if one wants to investigate the dynamics of the
system, since the algorithms used for constant pressure simulations
introduce an artificial dynamics which might lead to unphysical
results.  The step size we chose was 0.01 for the higher and 0.016 or
0.02 for the lower temperatures. For runs shorter or equal than
1.4$\cdot 10^5$ time units these step sizes are sufficiently small to
allow to neglect the drift in the total energy during the runs. This is
not the case for the long runs at the two lowest temperatures which had a
length of 3.0$\cdot 10^5$ and $4.0\cdot 10^5$ time units,
respectively, ($=1.5\cdot 10^7$ and $2.0 \cdot 10^7$ time steps).
During these runs the value of the total energy was reseted
periodically (about 30 times during the whole run) to its value at the
start of the run by rescaling the velocities of all the particles
appropriately. Since this interference with the dynamics is only very
weak it can be expected that the final result will essentially be
independent of it.

The temperatures we investigated are $T=5.0$, 3.0, 2.0, 1.4, 1.1, 0.85,
0.70, 0.632, 0.588, 0.549, 0.520, 0.500, 0.489, and 0.477. The total
number of molecules was 500 and in order to improve the statistics of
the results we averaged at each temperature over at least eight
independent runs.

\section{Results}
\label{sec4}

Before we start to present the results on how the ODOF freezes, it is
useful to investigate the dynamics of the TDOF, since this allows us to
estimate the temperature range in which the system is supercooled.
Therefore we computed from the mean square displacement of the center
of the molecules the translational diffusion constant $D$. MCT predicts
that, in the vicinity of the critical temperature $T_c$, the diffusion
constant will show a power-law [see Eq.~(\ref{eq9})]. Thus we fitted
our data for $D$ with such a law, using $T_c$ as a fit parameter. In
Fig.~\ref{fig1} we show the diffusion constant versus $T-T_c$ in a
double logarithmic plot. Also included is the fit with the power-law of
Eq.~(\ref{eq9}). We recognize that this fit is very good for a
surprisingly large range in $T$ and $D$. In particular this range is
significantly larger than the one found for the atomic Lennard-Jones
system~\cite{kob94,kob95b}. Since no analogous analysis was done for
the molecular systems studied in
Refs.~\cite{signorini90,sindzingre92,lewis94,kudchadkar95}, and the
range in $D$ explored in Ref.~\cite{sciortino96} was significantly
smaller than the one considered here, it is at the moment not clear
whether the fact that the power-law can be observed over such an
extended range in $D$ is a peculiarity of the present system or whether
it is a general feature of molecular systems.

 From the figure we also note that at the two lowest temperatures the
diffusion constant is significantly higher than it would be estimated
from the power-law which fits the data well at higher temperatures.
The reason for this are likely the so-called ``hopping
processes''~\cite{hopping_proc}, the contributions to the relaxation
that are not accounted for in the {\it idealized} version of
the MCT.  Thus it can be expected that for such low temperatures the
predictions of this idealized version of the theory are no longer
valid.

The value of the critical temperature $T_c$ is $0.475\pm0.005$ which
allows us to conclude that the TDOF of the system become very slow in
the vicinity of this temperature. This is also corroborated by our
investigation of the intermediate scattering function (coherent as well
as incoherent) for which we found that their $\alpha$-relaxation time
shows also a power-law with a critical temperature at
0.475~\cite{kammerer97}.  The critical exponent $\gamma$ of the power-law 
for $D$ is 2.20, which is in the range of values found for $\gamma$
in simple supercooled systems. In passing we mention that for this
molecular system the dynamics of the TDOF is qualitatively similar to
the one of simple liquids~\cite{kammerer97} and that therefore the
molecular character of the particles does not seem to affect the dynamics
of the TDOF significantly.

We now focus our attention to the ODOF. The first quantity we
investigate is the correlation function $C_1^{(s)}(t)$ which was
defined in Eq.~(\ref{eq1}). In Fig.~\ref{fig2}a we show this time
correlation function for all temperatures investigated. From this
figure we recognize that for high temperatures $C_1^{(s)}(t)$ decays
quickly to zero.  At intermediate temperatures it starts to show a weak
shoulder which, on lowering the temperature further, becomes more
pronounced. The time range for which this shoulder is observed
coincides with the one in which a plateau is observed in the
intermediate scattering function~\cite{kammerer97} and thus can be
identified with the $\beta$-relaxation regime.

 From this figure we also recognize that for intermediate and low
temperatures the shape of the curves in the $\alpha$-relaxation regime
seems to be almost independent of temperature, i.e. that the so-called
time temperature superposition principle (TTSP) holds [see
Eq.~(\ref{eq10})]. In order to investigate this point closer we plot in
Fig.~\ref{fig2}b the same curves versus the rescaled time
$t/\tau_1^{(s)}(T)$, where $\tau_1^{(s)}(T)$ is the $\alpha$-relaxation
time. We define $\tau_1^{(s)}(T)$ to be the time it takes the
correlation function to decay to $e^{-1}$ of its initial value. From
this figure we recognize that the TTSP does not hold very well, in that
the slope of the curves at $t/\tau_1^{(s)}(T)=1.0$ changes
significantly even at low temperatures. Thus we conclude that for this
type of correlation function of the ODOF the TTSP does not hold very
well, which is in contrast with the behavior of the TDOF of simple
liquids~\cite{kob94,kob95b} and of the present system~\cite{kammerer97}
as well as for $C_1^{(s)}(t)$ for the OTP model studied by Wahnst\"om
and Lewis~\cite{wahnstrom97}.  We also note that defining the
$\alpha$-relaxation time $\tau_1^{(s)}(T)$ in a different way, namely
by the time it takes the correlation function to decay to 10\% of its
initial value, does not change this conclusion, since the TTSP does not
hold with this second definition of $\tau_1^{(s)}$ either.

Next we investigate the time and temperature dependence of
$C_2^{(s)}(t)$, see Eq.~(\ref{eq1}), which is shown in
Fig.~\ref{fig3}.  From the upper panel of the figure we see that
$C_2^{(s)}(t)$ decays qualitatively similar to $C_1^{(s)}(t)$
(Fig.~\ref{fig2}a). For the former, however, the height of the shoulder
is lower than the one in $C_1^{(s)}(t)$, which is reasonable, since to
a first approximation this height is given by the value of the second
Legendre polynomial evaluated at the height of the shoulder in
$C_1^{(s)}$ [see Eq.~(\ref{eq1})]. Note that this height is a measure
for the corresponding nonergodicity parameter $f_l^c=\lim_{t\to
\infty} C_l(t)$ at $T_c$.

Since the area under the $\alpha$-peak and under the remaining part of
the spectrum $\chi_l''$ (i.e. the critical decay and the microscopic
peak) are related to $f_l^c$ and to $1-f_l^c$, respectively, our result
suggests that the minimum between the two peaks is less pronounced for
$\chi_1''$ than for $\chi_2''$, provided that the width of the
microscopic peak is about the same for $l=1$ and $l=2$.  This could be
the reason why the detection of this minimum is so difficult in
dielectric measurements, i.e. $l=1$, whereas it was readily found in
light scattering experiments.

 From Fig.~\ref{fig3}b we recognize that for this correlation function
the TTSP holds well for times larger than $\tau_2^{(s)}(T)$, the
$\alpha$-relaxation time for $C_2^{(s)}(t)$, but that for shorter times
quite significant discrepancies are observed, as it was the case for
$C_1^{(s)}$. Thus we come to the conclusion that the relaxation
behavior of $C_1^{(s)}$ and $C_2^{(s)}$ are qualitatively different.

Since the time dependence of $C_1^{(s)}$ seems to differ from the one
of $C_2^{(s)}$, we have also studied the one of $C_l^{(s)}$ for
$l=3,\ldots,6$, which, for the lowest temperature investigated, are
show in Fig.~\ref{fig3_new}b. From this figure we recognize that i) the
height of the plateau decreases with increasing $l$, which can be
rationalized by the same reasoning given above, ii) that with
increasing $l$ the correlators seem to show more and more a logarithmic
time dependence in the $\beta$-relaxation regime, and iii) that the
correlators for odd values of $l$ decay faster than the ones for even
values of $l$. This effect can be understood by taking into account
that at low temperatures the molecules make jump-like orientational
flips of $180^{\circ}$, described in more detail below, which lead to a
relaxation in $C_l^{(s)}$ if $l$ is odd, but do not affect the
correlators with even values of $l$. Furthermore we have found that
with increasing value of $l$ the TTSP holds better and
better~\cite{kammerer_thesis} thus showing that from a qualitative
point of view the correlators for the ODOF become more similar to the
ones of the TDOF.

In addition to the self parts, $C_l^{(s)}$ we have also determined
the time dependence of $C_1(t)$, one of the collective correlation
functions of the ODOF [see Eq.~(\ref{eq2})]. This correlation
function is shown in Fig.~\ref{fig3_n_n} for all temperatures
investigated. Although the noise in the data is significantly larger
than the one in $C_1^{(s)}$, as it is often the case for collective
quantities, we can clearly recognize that the time dependence of
$C_1$ is qualitatively similar to the one of $C_1^{(s)}$ and that the
TTSP, cf. Fig.~\ref{fig3_n_n}, does not seem to hold. Furthermore we
note that, e.g., at the lowest temperature and for $t=10^3$ time
units, $C_1^{(s)}(t)$ is about 25\% larger than $C_1(t)$, which
demonstrates that the cross terms in $C_1(t)$ should not be
neglected, at least not in the strongly supercooled regime.

 From Figs.~\ref{fig2}a,~\ref{fig3}a and \ref{fig3_n_n}a we recognize
that with decreasing temperature the relaxation of the ODOF slows down
dramatically. Thus it is interesting to investigate the temperature
dependence of the relaxation times. Since we have found that the
diffusion constant (Fig.~\ref{fig1}) as well as the $\alpha$-relaxation
times of the TDOF~\cite{kammerer97} show a power-law dependence on
temperature, with the same critical temperature $T_c$, we checked
whether also the $\alpha$-relaxation times $\tau_1^{(s)}(T)$,
$\tau_2^{(s)}(T)$ and $\tau_1(T)$ can be fitted with such a power-law
(with the same $T_c=0.475$). That this is indeed possible for about two
orders of magnitude in $\tau$, is demonstrated in Fig.~\ref{fig4} where
we show these quantities versus $T-T_c$ in a double logarithmic plot.
As it was the case for the diffusion constant $D$, the values of $\tau$
for the two lowest temperatures deviate from the power-laws, since also
here the relaxation is too fast, which is likely to be related to
hopping processes. Thus we conclude that these processes affect 
the ODOF also.

From Fig.~\ref{fig4} we also note that the fitted power-laws do not
extend to such high temperatures as they did in the case of the
diffusion constant. This is evidence that, similar to {\it simple}
liquids, the presence of a large temperature range for which such a
power-law is observed is rather the exception than the rule. The
critical exponents $\gamma$ of the three power-laws [see
Eq.~(\ref{eq8})], are 1.66, 2.42 and 1.52 for $\tau_1^{(s)}$,
$\tau_2^{(s)}$ and $\tau_1$.  Since the critical exponent for the
diffusion constant is 2.20, we thus find that the four critical
exponents are all different from each other, which disagrees with the
prediction of MCT for simple liquids. However, if we determine the
critical exponents for $C_l^{(s)}$ for $l=3,\ldots , 6$ we find the
values 2.25, 2.78, 2.55 and 2.80. These values have to be compared with
the critical exponent for the TDOF, which is around
2.6~\cite{kammerer97}, and thus quite close to the one of $C_l^{(s)}$
for the larger values of $l$. Thus this is more evidence that the
latter correlators behave qualitatively similar to the ones for the
TDOF.

We also mention that a power-law fit to $\tau_l^{(s)}$ and $\tau_1$,
with the critical temperature $T_c$ as a free parameter, leads to a
slightly improved fit. The optimal values of the critical temperature
were to within 2\% equal to 0.475, the value of $T_c$ from the
diffusion constant. Thus we find that the ODOF, measured by $C_l^{(s)}$
and $C_1$, would indeed freeze very close to $T_c=0.475$, if the
hopping processes were absent.

Furthermore we note that a fit of $\tau_1^{(s)}$ and $\tau_2^{(s)}$
with the popular Vogel-Fulcher law, $A\exp(B/(T-T_0))$, also works
remarkably well. In particular we find that this functional form is
able to fit the data well at {\it all} lower temperatures, i.e.  also
the data points at the two lowest temperatures, for which the power-law
failed to fit the data. Thus we conclude that if seen as a mere fitting
function the Vogel-Fulcher law gives the better fit than the
power-law.  However, the Vogel-Fulcher temperature $T_0$ was determined
to be 0.328 and 0.386 for $\tau_1^{(s)}$ and $\tau_2^{(s)}$,
respectively. Thus we find that the two temperatures differ by about
20\%, hence indicating that according to the Vogel-Fulcher fits there
is no unique temperature at which the system ceases to relax. Therefore
this sort of fit is, from a physical point of view, less appealing.

The $\alpha$-relaxation times $\tau_l^{(s)}(T)$ and $\tau_1(T)$ are
analogous quantities to the $\alpha$-relaxation time $\tau(T)$ of the
intermediate scattering function. Since in supercooled liquids the
temperature dependence of $\tau$ and of the diffusion constant can be
different (see, e.g., Refs.~\cite{kob95b,sciortino96}) it is
interesting to investigate also the {\it rotational} diffusion constant
$D_r$, and compare it with the temperature dependence of the relaxation
times $\tau_1^{(s)}$ and $\tau_2^{(s)}$.

As already mentioned in Sec.~\ref{sec2}, the calculation of $D_r$ is
numerically difficult when one uses a Green-Kubo relation. Instead it
is much simpler to compute $D_r$ from the Einstein relation given by
Eq.~(\ref{eq7}). In Fig.~\ref{fig5} we show the time dependence of the
mean square displacement of the angles $\vec{\phi}(t)$, i.e.  $\langle
(\Delta\vec{\phi}(t))^2\rangle=\langle
|\vec{\phi}_n(t)-\vec{\phi}_n(0)|^2 \rangle$, where $\vec{\phi}_n(t) -
\vec{\phi}_n(0)$ is defined in Eq.~(\ref{eq6}).

From this figure we recognize that, analogous to the mean square
displacement~\cite{kammerer97}, $\langle (\Delta
\vec{\phi}(t))^2\rangle$ shows at short times a power-law with exponent
2.0, which corresponds to the free rotational motion of the molecules.
For high temperature this type of motion crosses over directly into a
diffusional behavior, i.e.  $\langle(\Delta \vec{\phi}(t))^2\rangle$ is
given by a power-law with exponent 1.0.  This is not the case for the
low temperatures, where the time regimes of the free rotation and the
one of the diffusive behavior are separated by a time regime where the
increase of $\langle (\Delta \vec{\phi}(t))^2\rangle $ is slower than
diffusive. The time at which this subdiffusive behavior {\it starts} is
essentially the same as the one in which also the mean squared
displacement of the particles (MSD) starts to show a subdiffusive
behavior~\cite{kammerer97}. In contrast to this the time where
$\langle(\Delta \vec{\phi}^2(t))^2\rangle$ starts to show the diffusive
behavior is, at the lowest temperatures, significantly less (by about
1-2 decades) than the times where the MSD starts to show the diffusive
behavior. Thus, despite the qualitative similarity of the time
dependence of $\langle(\Delta \vec{\phi}(t))^2\rangle$ and the MSD
there are some distinct differences between the two quantities and thus
we conclude that the plateau-like region in $\langle(\Delta
\vec{\phi}(t))^2\rangle$ should {\it not} be identified with the
$\beta$-relaxation regime. We will come back to this point later.

From the time dependence of $\langle(\Delta \vec{\phi}^2(t))^2\rangle$
it is simple to compute $D_r$ [see Eq.~(\ref{eq7})]. Note that, because
$\langle (\Delta \vec{\phi}^2(t))^2\rangle$ reaches its diffusive limit
at shorter times than the MSD does, the rotational diffusion constant
can be calculated reliably from a relatively short run, an observation
of which we will make use of below.

In Fig.~\ref{fig6} we show the temperature dependence of $D_r$ in an
Arrhenius plot.  In order to facilitate the comparison between the
rotational and translational diffusion constant, we have included the
latter in the figure as well. (Note that we have multiplied $D$ by 15
in order to make $D$ and $D_r$ to coincide at high temperatures. Also
it should be remembered that $D$ and $D_r$ have different units.) We
see that for temperatures less than 2.0 the data (diamonds) can be
fitted well with an Arrhenius law (solid straight line). Furthermore
we recognize that the temperature dependence of $D_r$ follows
the one of $D$ for high temperatures but that when the system starts to
become supercooled, the curve for $D$ drops significantly below the one
for $D_r$. Thus we find that at high enough temperatures the ODOF and
the TDOF couple strongly enough to show the same temperature dependence
of $D_r$ and $D$, which is in agreement with the well-known
Stokes-Einstein- and Stokes-Einstein-Debye-relations. For lower
temperatures $D$ shows the power-law discussed in Fig.~\ref{fig1}, the
reason for which are likely the mode coupling effects. In contrast to
this, $D_r$ shows an Arrhenius law from which we can conclude that the
rotational motion of the molecule is an activated process. We will
study this process in more detail below.

We have also checked whether at low temperatures, i.e. $2.0 \geq T\geq
0.477$ the temperature dependence of $D_r$ can be fitted well by a
power-law and found that this is indeed possible with a critical
temperature around 0.38 (dashed line in Fig.~\ref{fig6}). This
temperature is significantly lower than the critical temperature $T_c$
we found for the diffusion constant, the intermediate scattering
function and the relaxation times of $C_l^{(s)}$ and $C_1$, which was
0.475. In order to discriminate between the two functional forms we
made use of the observation described above that the rotational
diffusion constant can be determined from a relatively short run
(compared to the $\alpha$-relaxation times of the TDOF), see
Fig.~\ref{fig5}. Thus we used the temperature dependence of the density
(obtained from our equilibrated runs at temperatures $T\geq 0.477$) to
estimate the volume of the system at $T=0.41$. We then set up the
volume of the system such that its density corresponded to this
extrapolated value and quenched the system to $T=0.41$. After allowing
the system to relax for $2.0\cdot 10^5$ time units we started to
measure the time dependence of $\langle (\Delta \vec{\phi})^2\rangle$
for three independent runs.  Note that this time is clearly not
sufficient to equilibrate the system with respect to its TDOF, but it
should at least allow the system to get reasonably close to its
equilibrium state at this temperature. The so obtained $\langle (\Delta
\vec{\phi}(t))^2 \rangle$ is included in Fig.~\ref{fig5} as well
(bottom curve). We see that even at this low temperature the diffusive
rotational behavior can be observed on the time scale of our
simulation.  Thus we could extract the corresponding value of $D_r$ and
have included it in Fig.~\ref{fig6} as well.  This data point lies
reasonably close to an extrapolation for the previously fitted
Arrhenius law and is completely off the curve for the power-law. (The
fact that this point lies slightly above this Arrhenius law can be
understood by taking into account that at this temperature the TDOF are
not quite relaxed.  Hence it can be expected that all the relaxation
times are smaller than they would be in an equilibrated sample and that
therefore the measured diffusion constant is too large~\cite{aging}.)
Thus we conclude that the rotational diffusion constant, as defined in
Eq.~(\ref{eq7}), follows an Arrhenius law even at very low temperatures
and that it is very unlikely that its temperature dependence is given
by a power-law.

In Fig.~\ref{fig4} we have seen that the relaxation times of the
orientational correlation functions show a power-law dependence on
temperature and that the critical temperature $T_c$ is very close to
the one of the translational diffusion constant or the one of the
intermediate scattering function. From Fig.~\ref{fig6} it is
recognized, however, that the rotational diffusion constant $D_r$ does
{\it not} show any exceptional temperature dependence in the vicinity
of $T_c$. At first view these two facts seem to contradict each other,
but as we will show now, this is not the case at all. In order for the
time correlation functions $C_1^{(s)}$ and $C_2^{(s)}$ to decay to zero
it is necessary that the orientation of the molecules changes by an
angle on the order of $180^{\circ}$ and $90^{\circ}$ in the case of
$C_1^{(s)}$ and $C_2^{(s)}$, respectively. In order to undergo such a
large change of orientation, a molecule has to wait until the cage
formed by the surrounding molecules breaks up.  The time for this to
happen is related to the relaxation time of the translational degrees
of freedom and thus we find that the relaxation times of $C_1^{(s)}$
and $C_2^{(s)}$ become very large when the temperature approaches
$T_c$.

For the rotational diffusion constant the situation is different. As
can be seen from Eqs.~(\ref{eq6}) and (\ref{eq7}), $D_r$ remains finite
as long as there is a possibility that $|\Delta \vec{\phi}(t)|^2$ increases
(linearly) with time. At low temperatures a molecule will not be able
to make large changes of its orientation but a small librational
(tumbling) motion is still possible (for the TDOF this corresponds to
the rattling of the particles in their cage), which was nicely
demonstrated by Renner {\it et al.} for a system of infinitely thin
rods on a cubic lattice~\cite{renner95}. It is not hard to see that
this librational motion gives rise to a diffusive movement of the
$z$-component of the vector $\vec{\phi}$ and hence to a finite value of
$D_r$ (here the $z$-axis is defined by the molecular axis of the
molecule at time zero).

In order to see this effect clearer we have investigated the
orientational dynamics of the molecules at $T=0.41$ in more detail.  At
this low temperature the orientation of most of the molecules stays for
a long time in the vicinity of the direction it was at time $t=0$.
Thus we determined the mean orientation of the $z$-axis of each
molecule by averaging its direction over a period of $4\cdot 10^3$ time
units.  Note that such a mean direction makes only sense if the
orientation of the molecule does not change significantly. Hence we
will restrict the following analysis to only those molecules for which
the maximum deviation from this mean axis was less than $45^{\circ}$.
In the considered time window this is the case for 74\% of the
molecules.  Having determined the mean $z$-axis, we chose a $x$ and $y$
axis perpendicular to the $z$ axis and computed the time dependence of
$\langle(\Delta \phi_{\alpha}(t))^2\rangle=\langle
|\phi_{\alpha}(t)-\phi_{\alpha}(0)|^2 \rangle$, with
$\alpha \in \{x,y,z\}$.  The time dependence of these three functions
are shown in Fig.~\ref{fig7}. We see that $\langle (\Delta
\phi_z(t))^2\rangle$ is indeed significantly larger than $\langle
(\Delta \phi_x(t))^2\rangle$ and $\langle(\Delta
\phi_y(t))^2$. This is in accordance with the picture put forward
above that the orientational diffusion of the molecules in the $z$
direction is much larger than the one in the $x$ and $y$ direction.  We
also recognize that the latter ones are not {\it completely} suppressed
which is likely to be due to the fact that the cage in which the
molecule sits is still slowly changing with time. It is important to
notice that similar arguments do not hold for the TDOF. The rattling
of the center of mass within a cage is isotropic an the average. It
is the direction of the molecular axis that breaks this isotropy on a
``mesoscopic'' time scale.

The fact that at low temperatures the molecules perform for a long time
a librational motion can also be demonstrated nicely by considering the
auto-correlation function $\Psi_2(t)$ of the angular velocity
$\vec{\omega}(t)$, see Eq.~(\ref{eq4}). As was illustrated by Renner {\it et
al.}, $\Psi_2(t)$ is expected to show a plateau with a height equal or
less than 0.25 if the motion of the molecule is of a librational
type~\cite{renner95}. In Fig.~\ref{fig7_n} we show the time dependence
of $\Psi_2$ for all temperatures investigated. From this figure we
recognize that the {\it short} time relaxation time of $\Psi_2$ {\it
decreases} with decreasing temperature and that at low temperatures the
correlation function shows indeed a plateau. 

A different way to study the orientational motion of the molecules is to
investigate the time dependence of the distribution function
$G(\theta,t)$, which is defined analogously to the self part of the van
Hove correlation function, i.e.  
\begin{equation} G(\theta,t) =
\frac{1}{N \sin\theta}\sum_{i=1}^N
\langle \delta[\theta-\cos^{-1}(\vec{u}_i(t)\cdot\vec{u}_i(0))]\rangle \quad,
\label{eq18} 
\end{equation}
where $\vec{u}_i(t)$ is the unit vector parallel to the axis of
molecule $i$ at time $t$.  In Fig.~\ref{fig8} we show $G(\theta,t)$ for
different temperatures.  Note that $G(\theta,t)$ is defined such that
for long times it approaches 1.0 for all values of $\theta$.
From Fig.~\ref{fig8}a we recognize that for $T=2.0$ this function
decays monotonically in $\theta$ for all times. This changes when the
temperature is decreased to $T=0.63$, Fig.~\ref{fig8}b, since then,
e.g., the curve for $t=77.7$ (bold dashed curve) shows a small
additional peak at $180^{\circ}$ which is separated from the main peak
at $0^{\circ}$ by a shallow minimum around $120^{\circ}$. This
additional peak stems from molecules which underwent a rotation of
$180^{\circ}$.  This feature becomes much more pronounced when the
temperature is decreased further to $T=0.477$ (Fig.~\ref{fig8}c).  The
mentioned minimum now exists for a large time range before it starts to
disappear on the time scale of the $\alpha$-relaxation. At even lower
temperature, $T=0.41$ (Fig.~\ref{fig8}d), the minimum does not show any
sign to fill up at all on the time scale of our simulation. However, we
see that the peak at $180^{\circ}$ is still observable, which means
that a significant fraction of the molecules made a flip of
$180^{\circ}$.

To study these $180^{\circ}$ jumps on a more microscopic level, we have
also investigated the time dependence of the angle $\theta$ of {\it
individual} molecules.  From such studies we found that at low
temperatures, i.e. $T=0.41$,  the $z$-axis of the molecules stays for a
long time in the vicinity of its orientation at $t=0$ and then
undergoes {\it relatively} quickly a flip of $180^{\circ}$ (see
Fig.~\ref{fig9} for three representative trajectories). The typical
time for this flip is around 50 time units, but also faster (see, e.g.,
Fig.\ref{fig9}b) as well as slower transitions can be observed.  This
transition time is relatively long compared to the time scale of a
(translational) vibration of a molecule in its cage which is on the
order of two time units. Therefore we conclude that such a
$180^{\circ}$ flip is not a fast process in which the molecule
overcomes one barrier in a quasi-ballistic way, but rather the sum of a
quick succession of smaller jumps. Finally we mention that the
molecules do not show these sort of little jumps neither before nor
after they undergo a $180^{\circ}$ flip which shows that these little
jumps are associated with the flips.

\section{Conclusions}
\label{sec5}

The main motivation of this paper has been to investigate the dynamics
of the orientational degrees of freedom in a supercooled molecular
liquid. This was done by means of a MD-simulation for a very simple
molecular system, a liquid of diatomic, rigid molecules.

The first question we addressed was how the translational and the
orientational degrees of freedom (TDOF and ODOF, respectively) slow
down if the temperature is decreased and the system becomes strongly
supercooled. Since the mode coupling theory (MCT) is presently the only
microscopic theory which predicts, in its idealized version, a glass
transition at a temperature $T_c$, we have checked the consistency of
our results with the predictions of the theory. 

In the $\alpha$-relaxation regime MCT predicts in the vicinity of $T_c$
a power-law behavior for the temperature dependence  of the
corresponding relaxation times and the diffusion constants [cf. Eqs.
(\ref{eq8}) and (\ref{eq9})]. We find that the translational diffusion
constant $D$ as well as the relaxation times $\tau$ for the
orientational correlators $C_l^{(s)}$ and $C_1$ can indeed be fitted by
a power-law and that the transition temperature $T_c$ is  $0.475
\pm 0.005$. This fit describes the data for $D$ and $\tau$ very well
for about four and two decades, respectively.  The corresponding
exponents are not universal, but vary between 1.52 and 2.45.

The next interesting result is that the rotational diffusion constant
$D_r$, as defined by Eq.~(\ref{eq7}), shows a significantly different
temperature dependence than the quantities just discussed.  For high
temperatures $D_r$ and $D$ are proportional to each other, in
accordance with the hydrodynamic Stokes-Einstein and
Stokes-Einstein-Debye relations. However, below a temperature
$T^*\approx 1.4$, which is far above $T_c$, $D$ is described well by
the mentioned power-law, whereas $D_r$ shows an Arrhenius dependence.
We find that $T^*$ is the temperature at which the cage-effect starts
to become important, i.e.  the system begins to be supercooled. This
can be inferred from the fact that at $T^*$ $\langle \Delta
r^{2}\rangle$ as well as $\langle (\Delta \vec{\phi})^2\rangle$ start
to show anomalous diffusion behavior at intermediate times. Thus we
conclude that the breakdown of this aspect of 
the hydrodynamic equations and the onset
of the cage-effect in supercooled liquids occur at the same
temperature. We stress, however, that the different temperature
dependence of $D$ and $D_r$ below $T^*$ is {\it not} related to the
similar phenomenon observed in experiments~\cite{fujara92,roessler}.
In these experiments the relaxation of the orientational vector
$\vec{u}_{n}(t)$ of  the $n$-th molecule is studied. The measured
quantity corresponds to the correlator $C_l^{(s)}(t)$ [see
Eq.~(\ref{eq1})]. In most theoretical approaches the rotational
dynamics is described by a Smoluchowski equation in which it is assumed
that the angular velocities can be eliminated adiabatically. This
crucial assumption, and the subsequent linearization, leads to an {\it
exponential} relaxation with a $\alpha$-relaxation time $\tau_l^{(s)}$
which is proportional to $1/D_r l(l+1)$~\cite{bagchi}.  Using this
relationship, $D_r$ can then be deduced.  Our results at lower
temperatures are not consistent with this theoretical result, because
i) $D_r(T)$ is {\it not} proportional to $ [\tau_l^{(s)}(T)]^{-1} $ and ii)
$\tau_1^{(s)}(T) / \tau_2^{(s)}(T) \neq 3$. The reason for this is
the nonexponential relaxation we have found for which the
proportionality of $D_r$ and $(\tau_l^{(s)})^{-1}$ and the relation
$\tau_1^{(s)}/\tau_2^{(s)}=3$ is not granted.

In the vicinity of $T_c$ the data for $D$ and $\tau_l^{(s)}$ deviate
from the power-law observed at intermediate temperatures.  (The same
deviations are found for the $\alpha$-relaxation times of the coherent
and incoherent intermediate scattering function of the center of the
molecules~\cite{kammerer97}). This is likely due to the hopping
processes which restore ergodicity even at low temperatures. If hopping
of the center of mass of the molecules becomes important at lower
temperatures, this should be seen in the self part of the van Hove
correlation function $G_s(r,t)$~(see, e.g.,
Ref.~\cite{roux_wahnstrom}).  Surprisingly, even at the lowest
temperature, $T=0.477$, $G_s(r,t)$ does not show any sign of a second
(smaller) peak at a distance $r\approx 1$, which corresponds to the
mean distance between two neighboring molecules~\cite{kammerer97}.
That $D(T)$ nevertheless deviates from a power-law at low $T$ may be
explained by the jump-like reorientations of the ODOF which we have
identified in the distribution function $G(\theta,t)$ as
$180^{\circ}$-flips. These jumps, which were also observed in earlier
MD-simulations~\cite{signorini90,lewis94}, may lead to a local
``melting'' of the neighborhood of a molecule which just jumped,
whereby allowing a translational diffusion without the molecule having
to jump over the walls of its cage.  The rotational diffusion measured
by $D_r$ is an activated process, at least for lower temperatures. Its
$T$-dependence can be described by an Arrhenius law even for a
temperature $T=0.41$ which is far below $T_c$. This Arrhenius
dependence is somewhat reminiscent to the temperature dependence of the
Johari-Goldstein $\beta$-peak in the dynamic
susceptibility~\cite{johari71}. Since we have shown that the Arrhenius
dependence of $D_r$ is related to the librational motion of the
molecules, one thus might speculate whether the dynamics leading to the
$\beta$-peak is indeed related to such librations. However, in order to
decide this one would have to investigate the equilibrium dynamics of
the systems at significantly lower temperatures than it is presently
possible.

Regarding the time dependence of the various correlators at low
temperatures we have found a two-step relaxation process, with a
strongly nonexponential behavior, for all of them. This is in agreement
with the results in Refs.~\cite{signorini90,lewis94,sciortino96} and
also with MCT.  However, in contrast to the prediction of MCT, the
time-temperature superposition principle does not seem to work very
well for the orientational correlators investigated here, although it
does so for $C_l^{(s)}$ with $l\geq 3$ and for the correlators of the
TDOF. This shows that, despite the fact that the temperature dependence
of the relaxation times of the TDOF and the one of the ODOF are very
similar, the relaxation dynamics of the two types of correlation
functions is, if $l=1$ or $l=2$, qualitatively different.  Finally we
mention that at low temperatures $C_1^{(s)}(t)$ and $C_1(t)$ differ
from each other in the $\alpha$-relaxation regime by about 25\%, which
demonstrates that the cross-terms in $C_1(t)$ should not be neglected.

Acknowledgements: We thank the DFG, through SFB 262, for financial
support. Part of this work was done on the computer facilities of the
Regionales Rechenzentrum Kaisers\-lautern.

\newpage

\begin{figure}
\caption{Self-diffusion constant $D$ versus $T-T_c$. The solid line
is a fit with a power-law with the exponent 2.20. The dashed line is
a guide to the eye.
\protect\label{fig1}}
\vspace*{5mm}
\par
\caption{Time dependence of $C_1^{(s)}$ [see Eq.~(2)] for all
temperatures investigated. a) versus time~$t$ b) versus rescaled time
$t/\tau_1^{(s)}(T)$, where $\tau_1^{(s)}$ is the $\alpha$-relaxation time.
\protect\label{fig2}}
\vspace*{5mm}
\par
\caption{Time dependence of $C_2^{(s)}$ [see Eq.~(2)] for all
temperatures investigated. a) versus time $t$ b) versus rescaled time
$t/\tau_1^{(s)}(T)$, where $\tau_1^{(s)}$ is the $\alpha$-relaxation time.
\protect\label{fig3}}
\vspace*{5mm}
\par
\caption{Time dependence of $C_l^{(s)}$ for $l=1,\ldots,6$ for
$T=0.477$, the lowest temperature investigated.
\protect\label{fig3_new}}
\vspace*{5mm}
\par
\caption{Time dependence of $C_1$ [see Eq.~(1)] for all
temperatures investigated. a) versus time~$t$ b) versus rescaled time
$t/\tau(T)$, where $\tau$ is the $\alpha$-relaxation time.
\protect\label{fig3_n_n}}
\vspace*{5mm}
\par
\caption{$\alpha$-relaxation time $\tau_1^{(s)}$ (squares),
$\tau_2^{(s)}$ (diamonds) and $\tau_1$
(circles) versus temperature. Solid lines: fits with power-law.
The dashed lines are guides to the eye.
\protect\label{fig4}}
\vspace*{5mm}
\par
\caption{Time dependence of $\langle (\Delta \vec{\phi}(t))^2\rangle$ 
[see~Eq.~(6)] for all temperatures investigated.
\protect\label{fig5}}
\vspace*{5mm}
\par
\caption{Temperature dependence of the rotational diffusion constant
$D_r$ (diamonds) and of the translational diffusion constant $D$
(circles). $D$ is multiplied by 15 so that the two curves coincide at
high temperatures. The straight solid line is an Arrhenius behavior
and the dashed line is a power-law with a critical temperature 0.38.
The dotted lines are guides to the eye.
\protect\label{fig6}}
\vspace*{5mm}
\par
\caption{Time dependence of $\langle (\Delta \phi_x)^2\rangle$,
$\langle (\Delta \phi_y)^2\rangle$ and $\langle (\Delta
\phi_z)^2\rangle$ and their sum $\langle (\Delta \vec{\phi})^2\rangle$
[see~Eq.~(6)] for $T=0.41$. See text for details.
\protect\label{fig7}}
\vspace*{5mm}
\par
\caption{Time dependence of the auto-correlation function $\Psi_2(t)$ of the
angular velocities [see~Eq.~(4)] for all temperatures investigated.
\protect\label{fig7_n}}
\vspace*{5mm}
\par

\caption{The function $G(\theta,t)$ for different times and
temperatures. Consecutive curves are spaced by about a factor of two in
time. The first curves (bold lines) correspond to a time of 
approximately 0.64. The
inset show the same curves on an expanded scale. a) $T=2.0$, b)
$T=0.63$, c) $T=0.477$, d) $T=0.41$.
\protect\label{fig8}}
\vspace*{5mm}
\par
\caption{Time dependence of the angle $\theta$ of an individual
molecule in the time rage where the molecule makes a $180^{\circ}$
flip. Three representative curves are shown. $T=0.41$.
\protect\label{fig9}}
\vspace*{5mm}
\par
\end{figure}


\begin{references}
\bibitem[\dag]{wkob}
Electronic mail: kob@moses.physik.uni-mainz.de\\
http://www.cond-mat.physik.uni-mainz.de/\~{ }kob/home\_kob.html
%
\bibitem[*]{rschilling}
Electronic mail: schillin@einstein.physik.uni-mainz.de
%
\bibitem{gotze91}
W. G\"otze in {\it Liquids, freezing and the glass transition}, Eds. J. P.
Hansen, D. Levesque and J. Zinn-Justin (North-Holland, Amsterdam,
1991).
%
\bibitem{gotze92}
W. G\"otze and L. Sj\"ogren, Rep. Prog. Phys. {\bf 55}, 241 (1992).
%
\bibitem{schilling94}
R. Schilling in {\it Disorder Effects on Relaxation Processes}, Eds. R.
Richert and A. Blumen (Springer, Berlin, 1994).
%
\bibitem{kob97}
W. Kob in {\it Experimental and Theoretical Approaches to
Supercooled Liquids: Advances and Novel Applications}, Eds.: J.
Fourkas, D. Kivelson, U. Mohanty, and K. Nelson (ACS Books, 
Washington, 1997).
%
\bibitem{yip95}
Theme Issue on Relaxation Kinetics in Supercooled Liquids-Mode Coupling
Theory and its Experimental Tests; Ed. S. Yip. Volume {\bf 24}, No.
6-8 (1995) of {\it Transport Theory and Statistical Physics}.
%
\bibitem{li92}
G. Li, M. Du, A. Sakai and H. Z. Cummins, Phys. Rev. {\bf A 46}, 
3343 (1992).
%
\bibitem{dixon90}
P. K. Dixon, L. Wu, S. R. Nagel, B. D. Williams and J. P. Carini, 
Phys. Rev. {\bf B 42}, 8179 (1990);
%
P. K. Dixon, Phys. Rev. {\bf B 42}, 8179 (1990).
%
\bibitem{lunkenheimer96}
P. Lunkenheimer, A. Pimenow, B. Schiener,  R. B\"ohmer and A. 
Loidl, Europhys. Lett. {\bf 33}, 611 (1996).
%
\bibitem{lunkenheimer962}
P. Lunkenheimer, A. Pimenow, M. Dressel, Y. G. Gonchunov, 
R. B\"ohmer and A. Loidl, Phys. Rev. Lett.  {\bf 77}, 318 (1996).
%
\bibitem{franosch97a}
T. Franosch, M. Fuchs, W. G\"otze, M.R. Mayr, and A.P. Singh,
Phys. Rev. E {\bf 55}, xxxx, (1997).
%
\bibitem{lunkenheimer963}
P. Lunkenheimer,  A. Pimenow, M. Dressel, Y. G. Gonchunov, U.
Schneider, B. Schiener R. B\"ohmer and A. Loidl, Proc. MRS Fall
Meeting, Boston 1996.
%
\bibitem{kob_stauf}
W. Kob, p.1 in Vol. III of {\it Annual Reviews of Computational Physics},
Ed.: D. Stauffer (World Scientific, Singapore, 1995).
%
\bibitem{roux_wahnstrom}
J.-N. Roux, J.-L. Barrat and J.-P. Hansen, J. Phys.: Cond. Matter
{\bf 1}, 7171 (1989);
%
G. Wahnstr\"om, Phys. Rev. A {\bf 44}, 3752 (1991).
%
\bibitem{kob94}
W.~Kob and H.~C. Andersen, Phys. Rev. Lett. {\bf 73}, 1376
(1994).
%
\bibitem{kob95b}
W.~Kob and H.~C. Andersen, Phys. Rev. E {\bf 51}, 4626 (1995); 
%
{\it ibid.} {\bf 52}, 4134 (1995).
%
\bibitem{signorini90}
G. F. Signorini, J.-L. Barrat and M. L. Klein, J. Chem. Phys. 
{\bf 92}, 1294 (1990).
%
\bibitem{sindzingre92}
P. Sindzingre, M. L. Klein, J. Chem. Phys. {\bf 96}, 4681 (1992).
%
\bibitem{lewis94}
L. J. Lewis and G. Wahnstr\"om, Phys. Rev {\bf E 50}, 3865 (1994).
%
\bibitem{wahnstrom97}
G. Wahnstr\"om and L. J. Lewis, Prog. Theor. Phys. (in press).
%
\bibitem{kudchadkar95}
S. R. Kudchadkar and J. M. Wiest, J. Chem. Phys. {\bf 103}, 8566 (1995).
%
\bibitem{sciortino96}
P. Gallo, F. Sciortino, P. Tartaglia and S.-H. Chen, 
Phys. Rev. Lett. {\bf 76}, 2730 (1996);
%
F. Sciortino, P. Gallo, P. Tartaglia and S. H. Chen, Phys. Rev. E
{\bf 54}, 6331 (1996).
%
\bibitem{sciortino_mrs}
F. Sciortino, P. Tartaglia and P. Gallo, Proc. MRS Fall
Meeting, Boston 1996.
%
\bibitem{kammerer97}
S. K\"ammerer, W. Kob and R. Schilling, (unpublished).
%
\bibitem{ma97}
W. J. Ma and S. K. Lai, Phys. Rev. E {\bf 55}, 2026 (1997).
%
\bibitem{franosch97b}
T. Franosch, W. G\"otze and A. P. Singh, (unpublished).
%
\bibitem{scheidsteger97}
T. Scheidsteger and R. Schilling, (unpublished).
%
\bibitem{schmitz97}
R. Schmitz, (unpublished).
%
\bibitem{gotze85}
W. G\"otze, Z. Physik {\bf B 60}, 195 (1985).
%
\bibitem{cummins96}
H. Z. Cummins, G. Li, W. Du, R. Pick, and C. Dreyfus, Phys. Rev. E
{\bf 53}, 896 (1996).
%
\bibitem{renner95}
C. Renner, H. L\"owen and J. L. Barrat, Phys. Rev. {\bf E 52}, 5091 
(1995).
%
\bibitem{allen87}
M. P. Allen and D. Frenkel, Phys. Rev. Lett. {\bf 58}, 1748 (1987).
%
\bibitem{andersen81}
H. C. Andersen, J. Chem. Phys. {\bf 72}, 2384 (1980).
%
\bibitem{andersen83}
H.C. Andersen, J. Comp. Phys. {\bf 52}, 24 (1983).
%
\bibitem{hopping_proc}
W. G\"otze and L. Sj\"ogren, Z. Phys. {\bf B 65}, 415 (1987).
%
\bibitem{kammerer_thesis}
S. K\"ammerer, PhD Thesis, University of Mainz, 1997.
%
\bibitem{aging}
See, e.g., J. Baschnagel, Phys. Rev. B {\bf 49}, 135 (1994); J.-P.
Bouchaud, L. Cugliandolo, J. Kurchan, and M. M\'ezard, Physica A {\bf
226}, 243 (1996); W. Kob and J.-L. Barrat cond-mat/9704006.
%
\bibitem{fujara92}
F. Fujara, B. Geil, H. Sillescu, and G. Fleischer, Z. Physik B {\bf 88}, 
195 (1992).
%
\bibitem{roessler}
E. R\"ossler, Phys. Rev. Lett. {\bf 65}, 1595 (1990).
%
\bibitem{bagchi}
B. Bagchi and A. Chandra, Adv. Chem. Phys. {\bf 80}, 1 (1991).
%
\bibitem{johari71}
G. P. Johari and M. Goldstein, J. Chem. Phys. {\bf 53}, 2372 (1971); 
{\it ibid.} {\bf 55}, 4245 (1971).
%
\end{references}
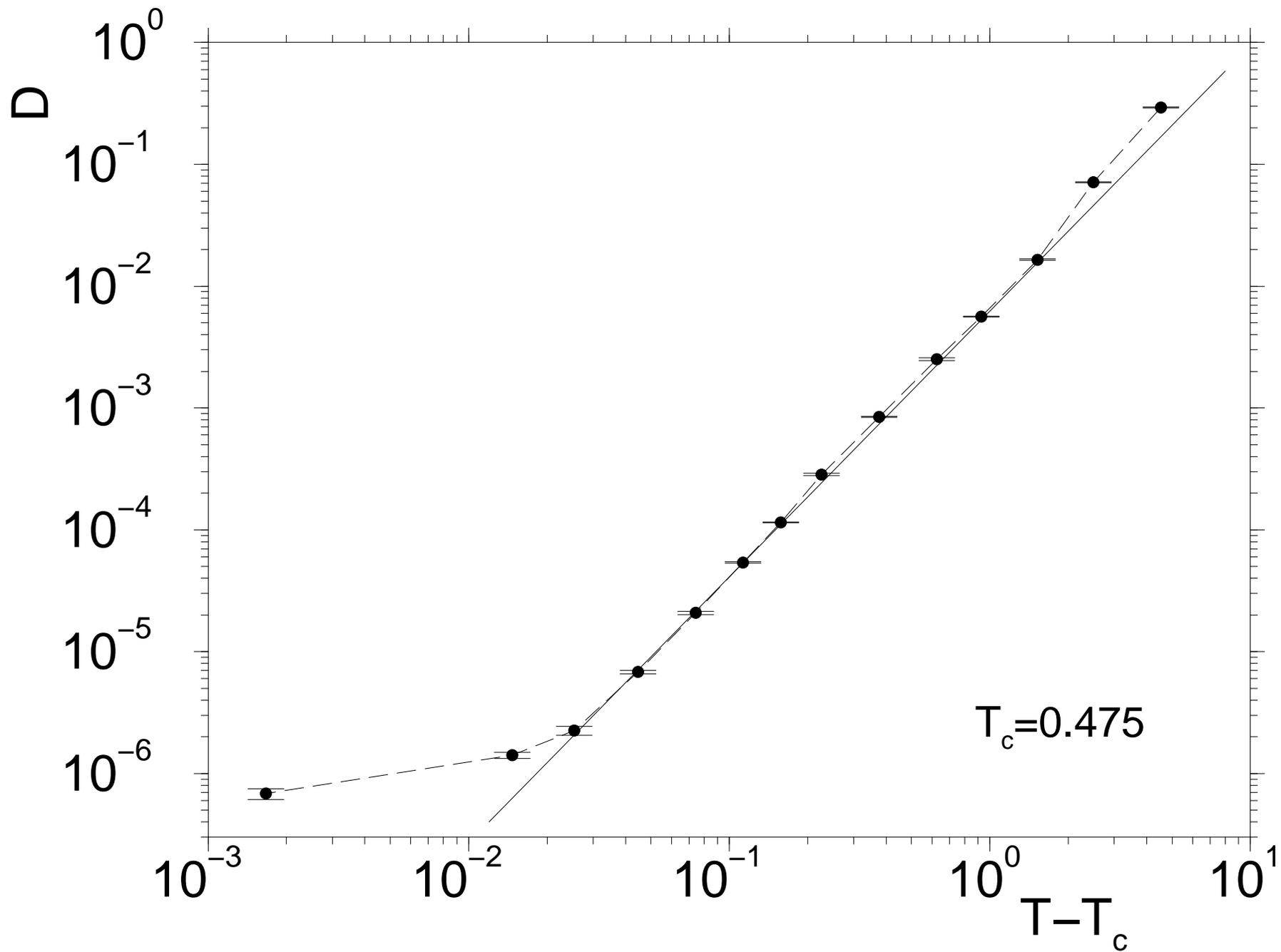
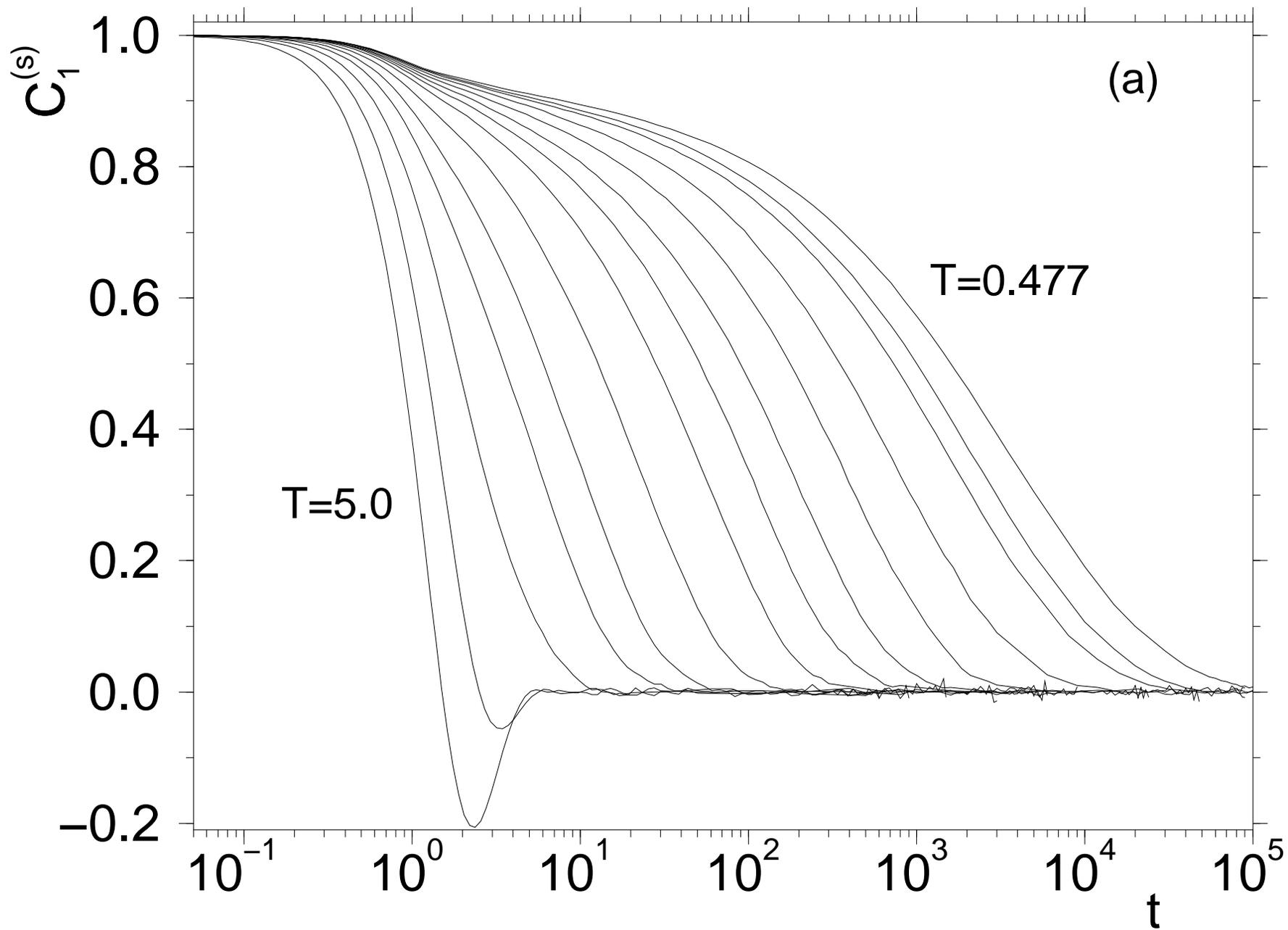
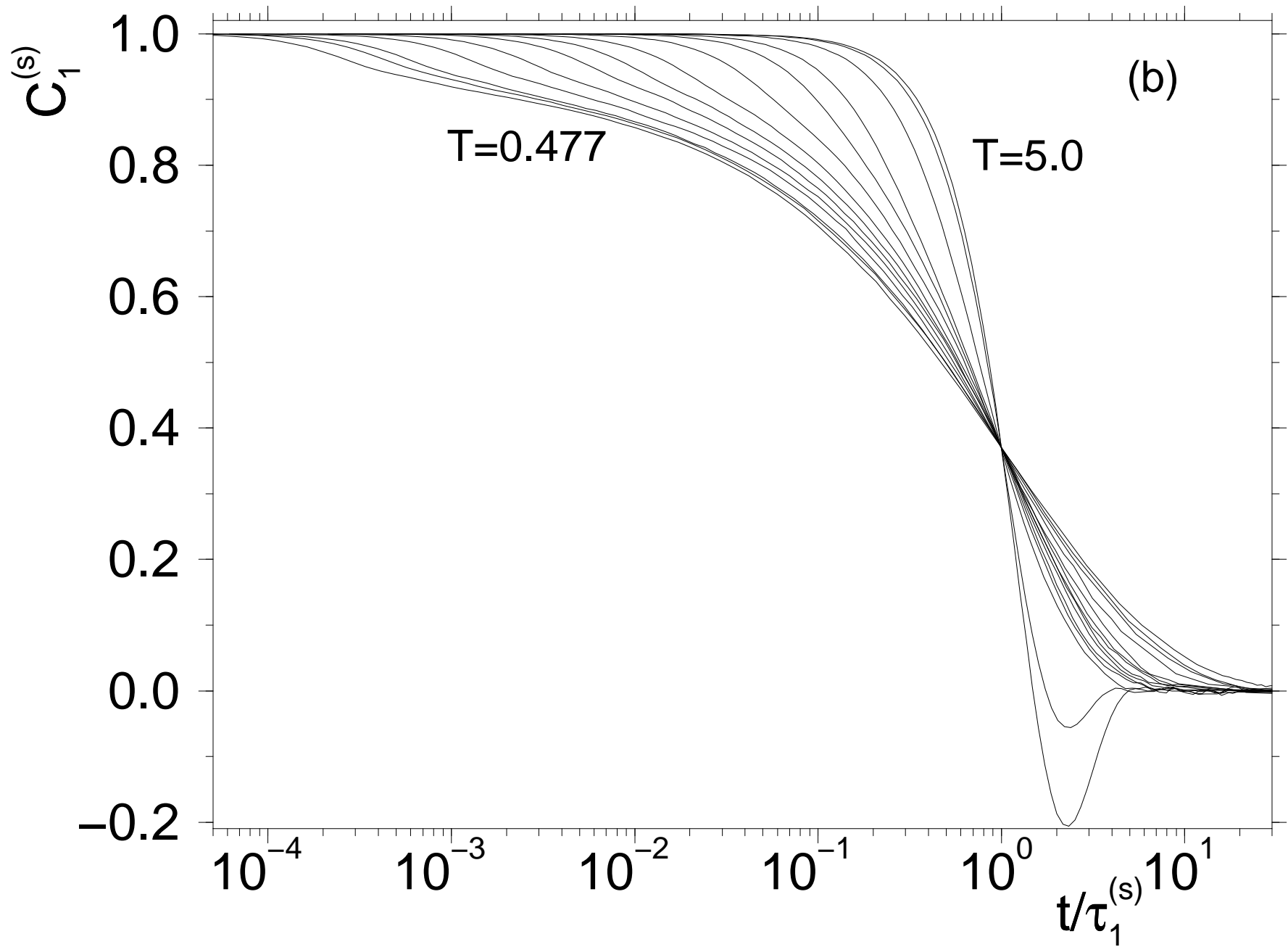
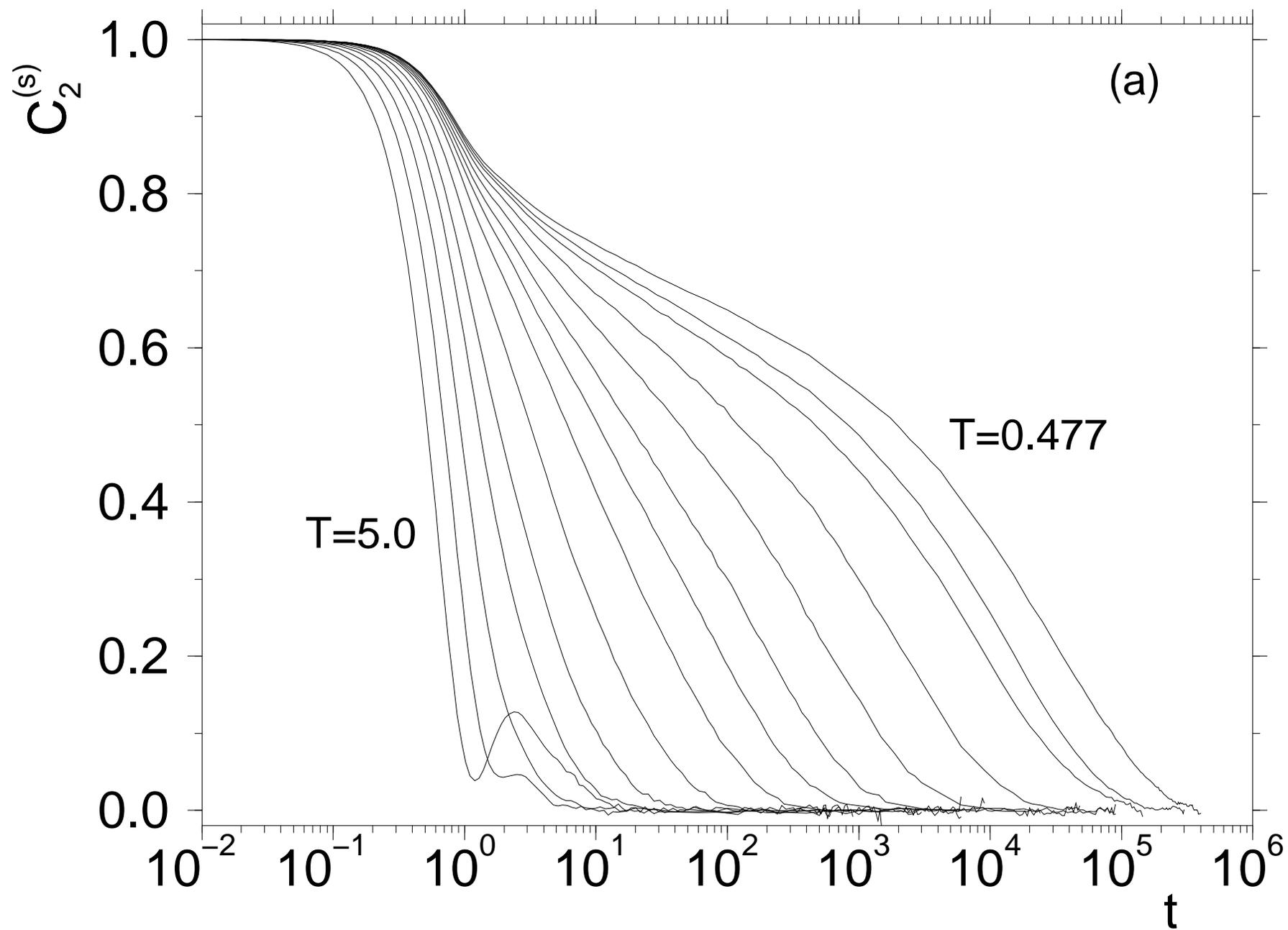
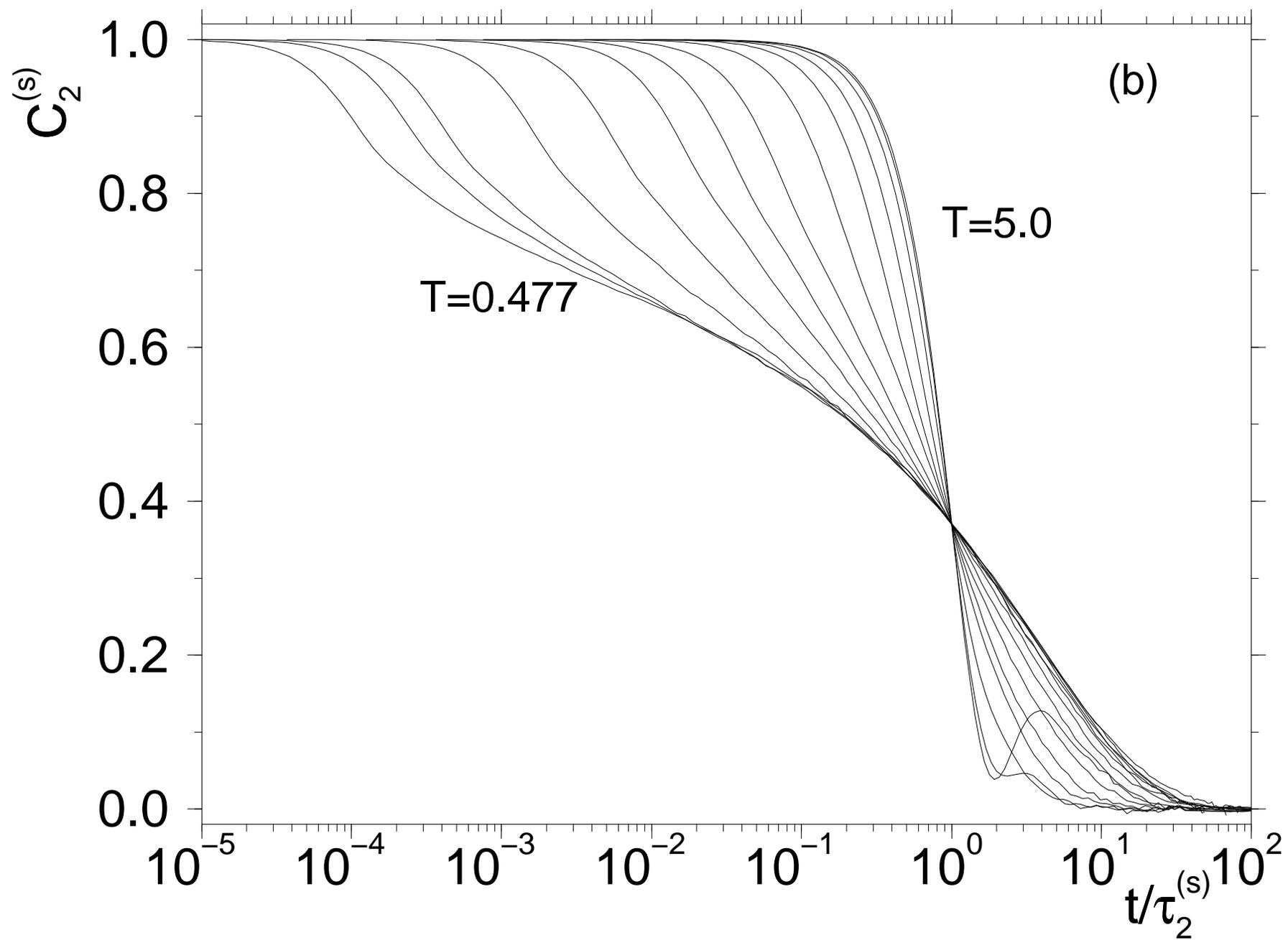
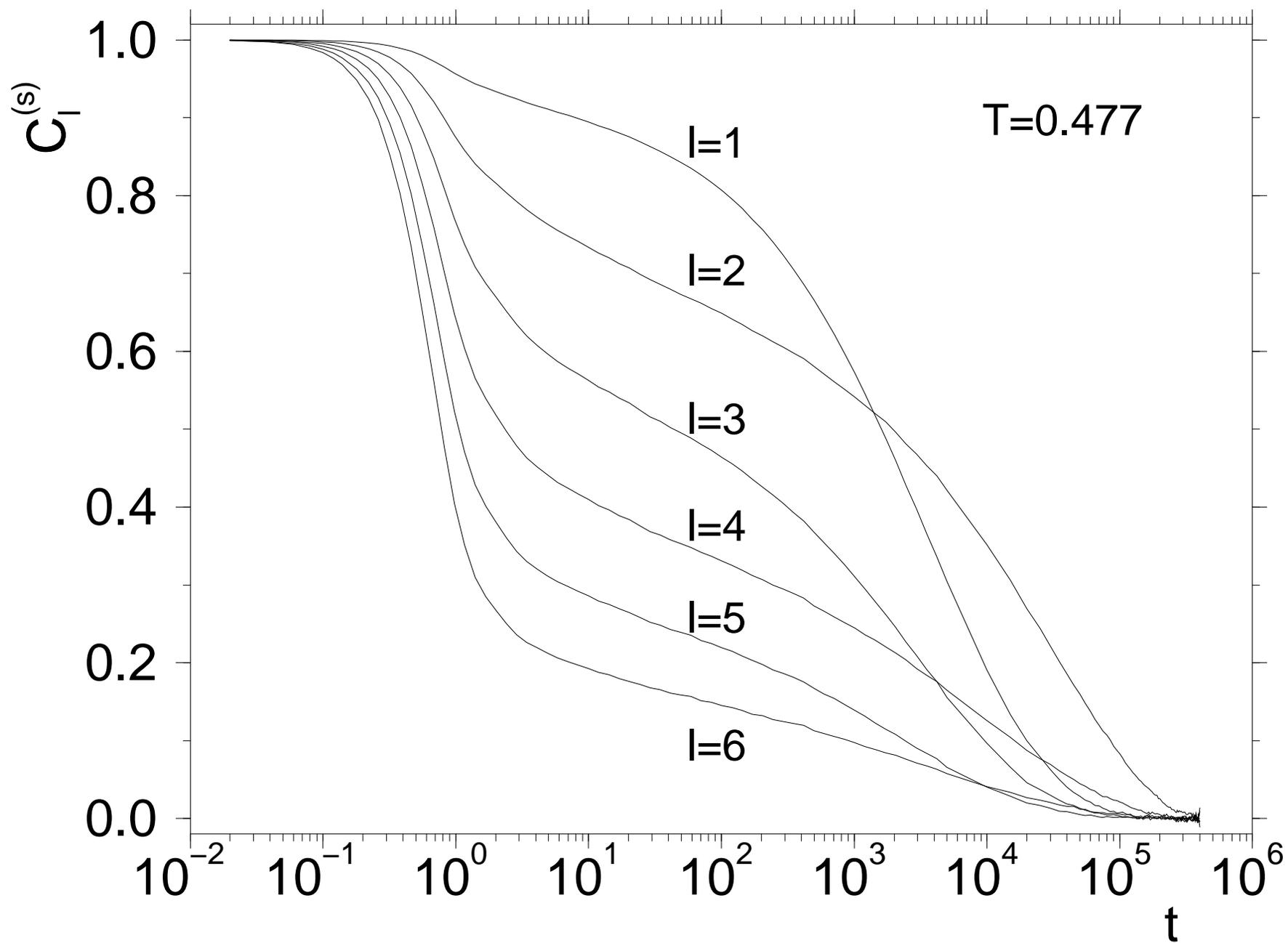
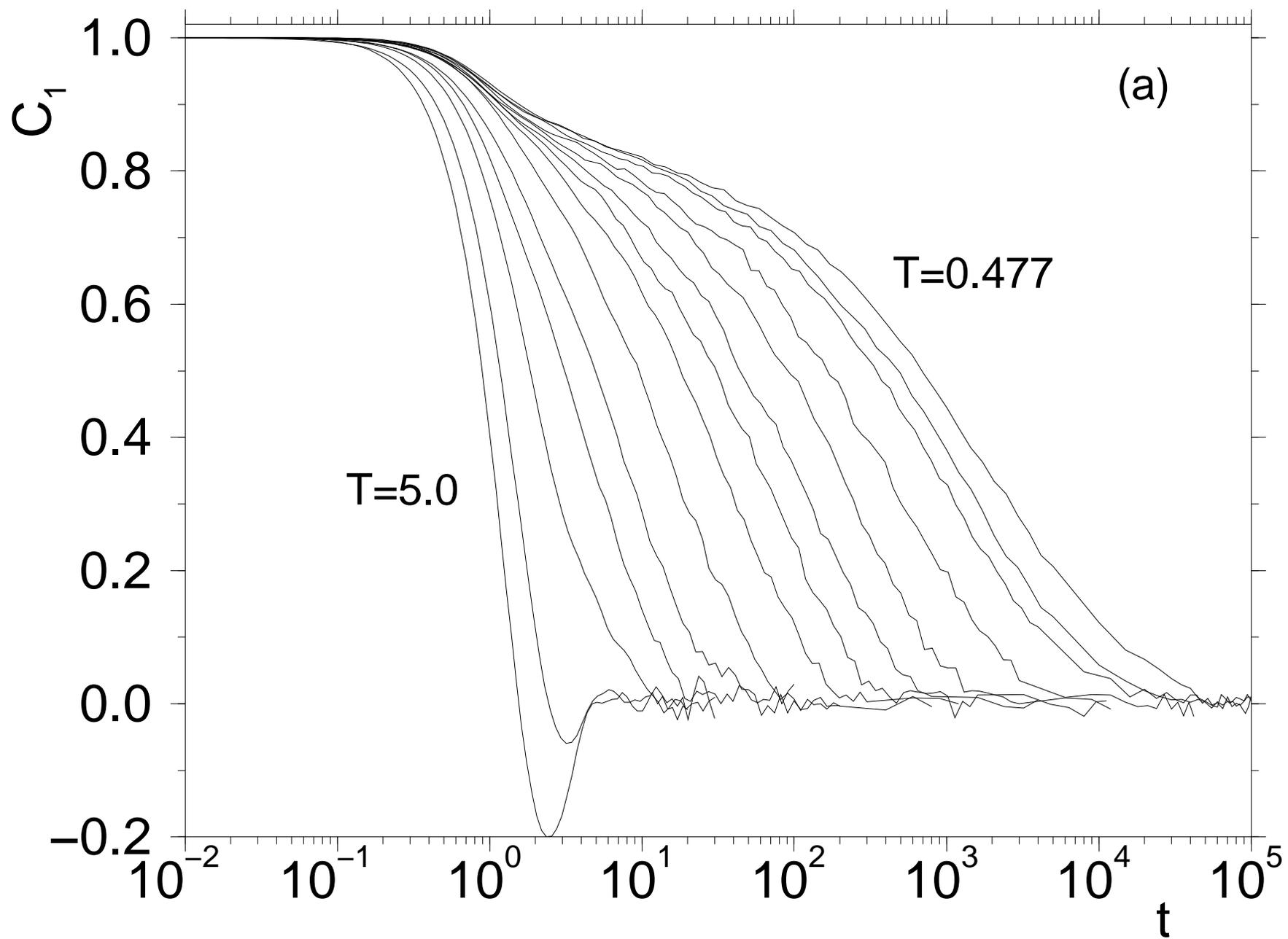
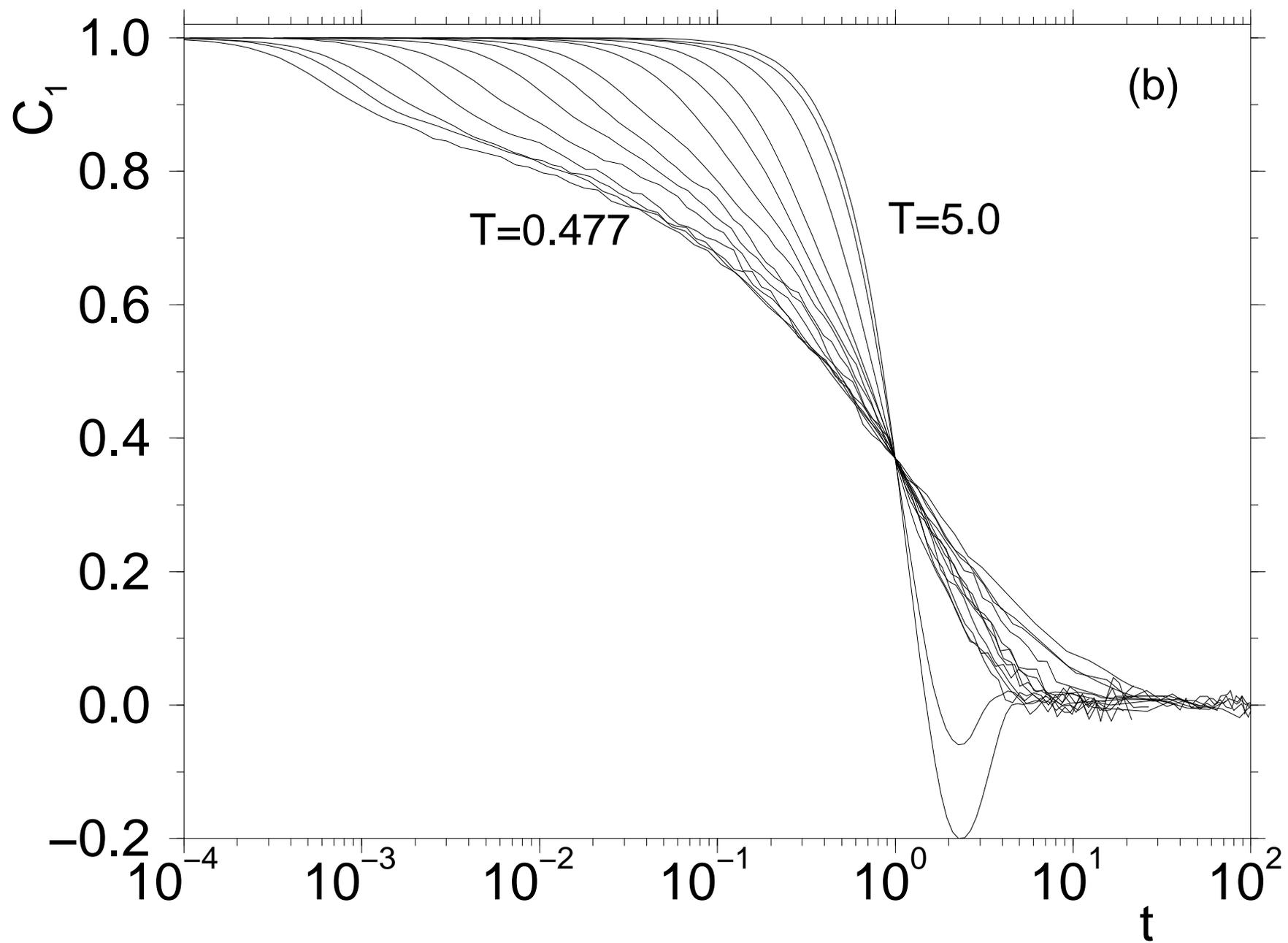
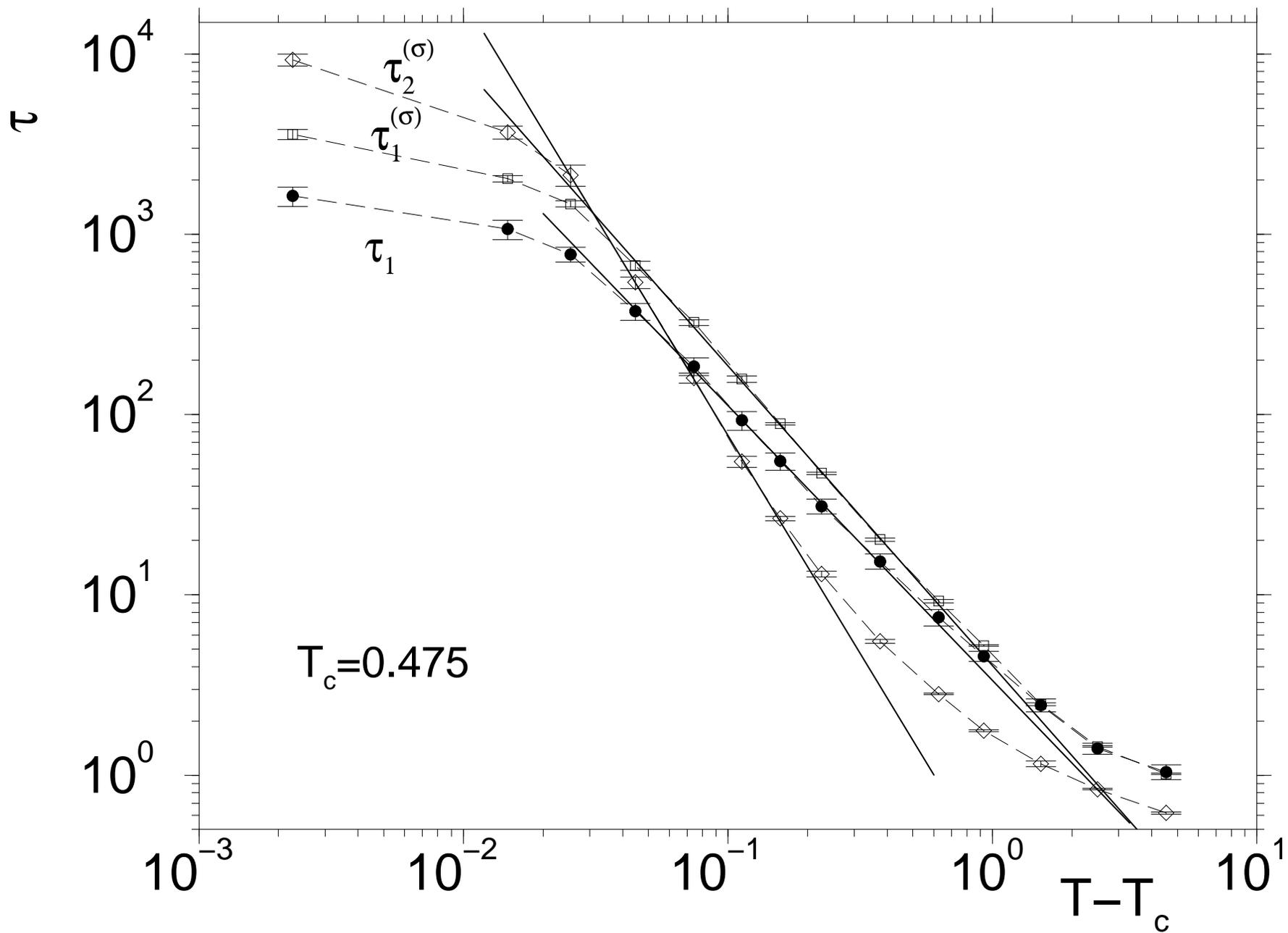
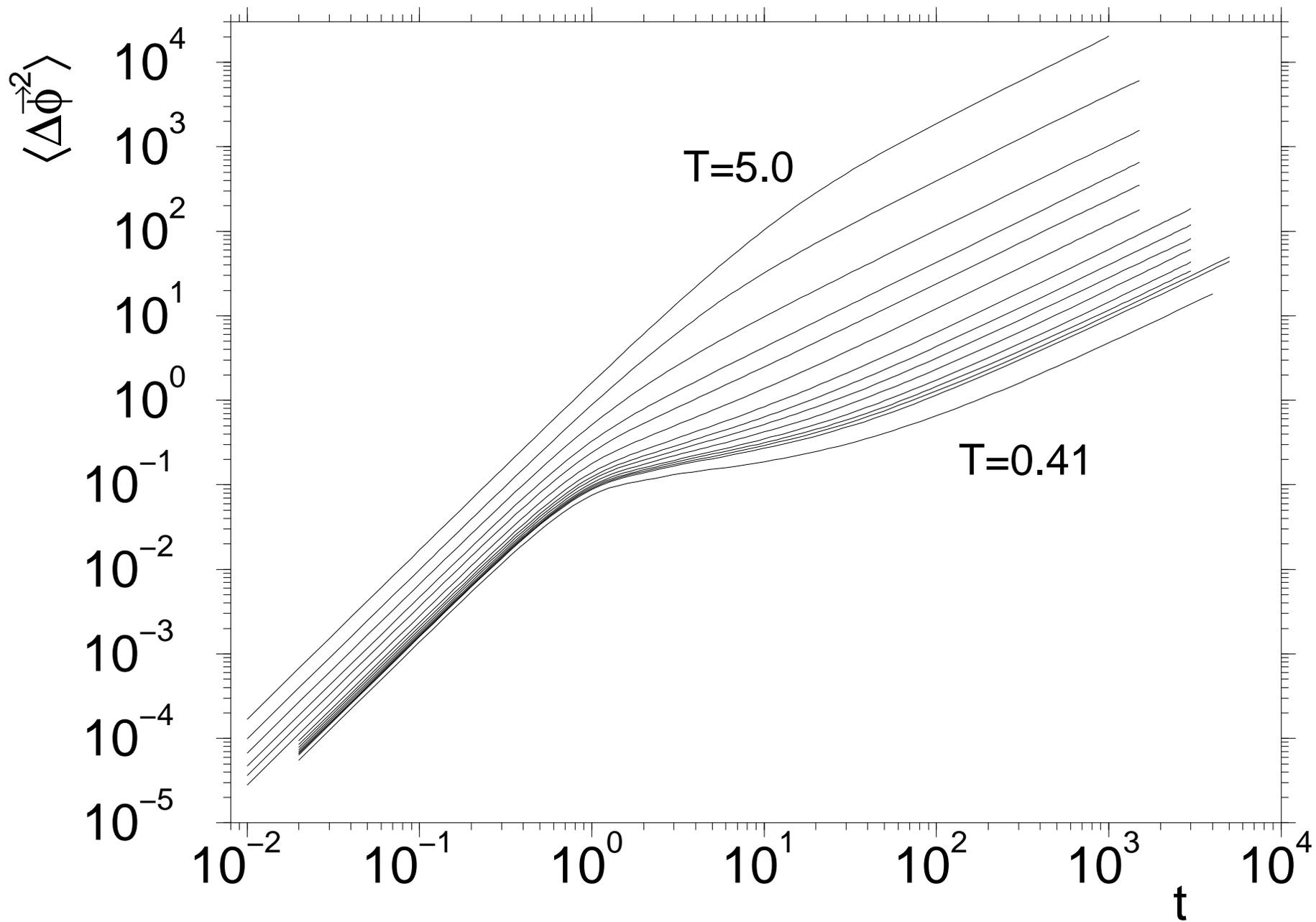
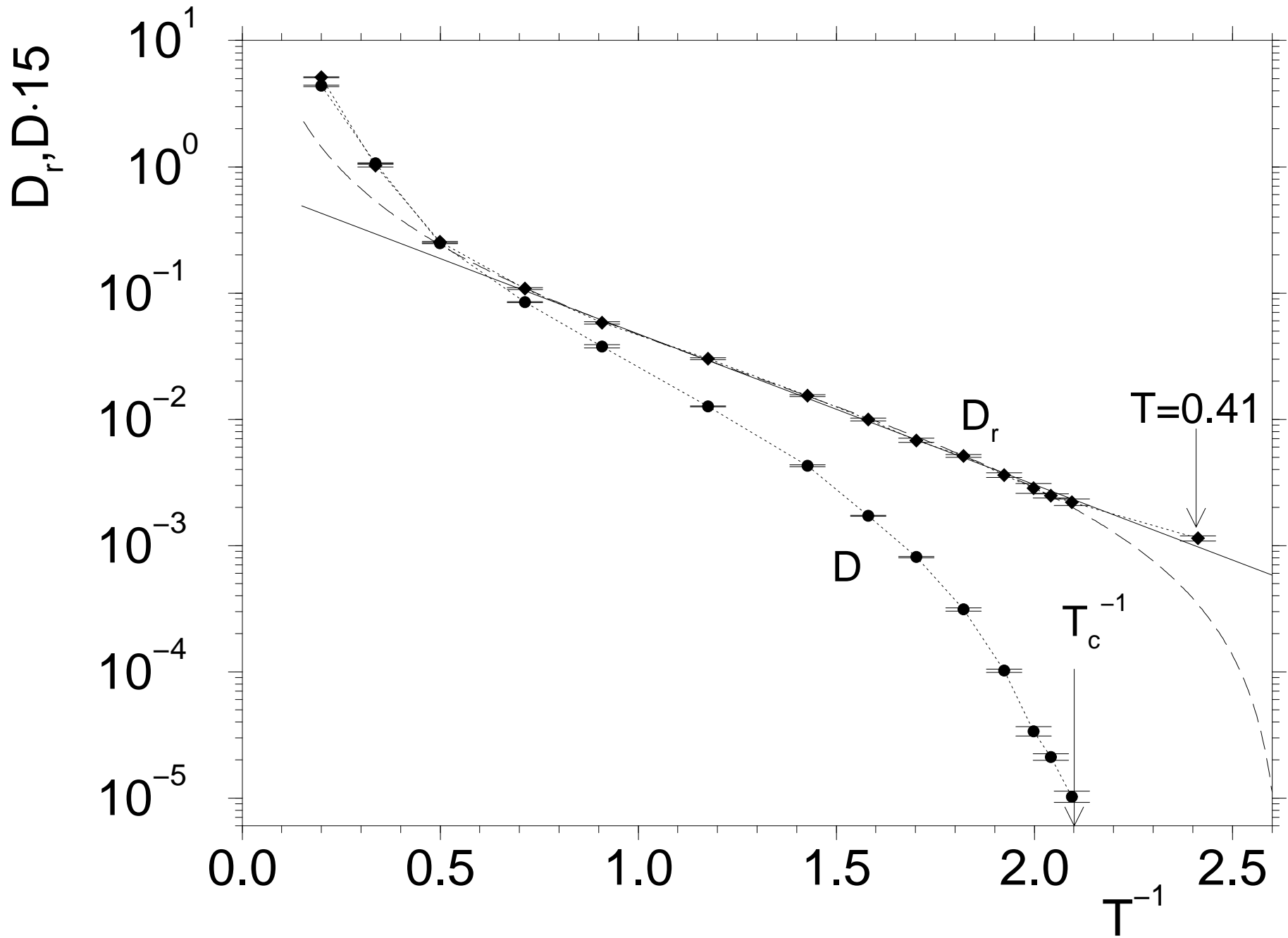
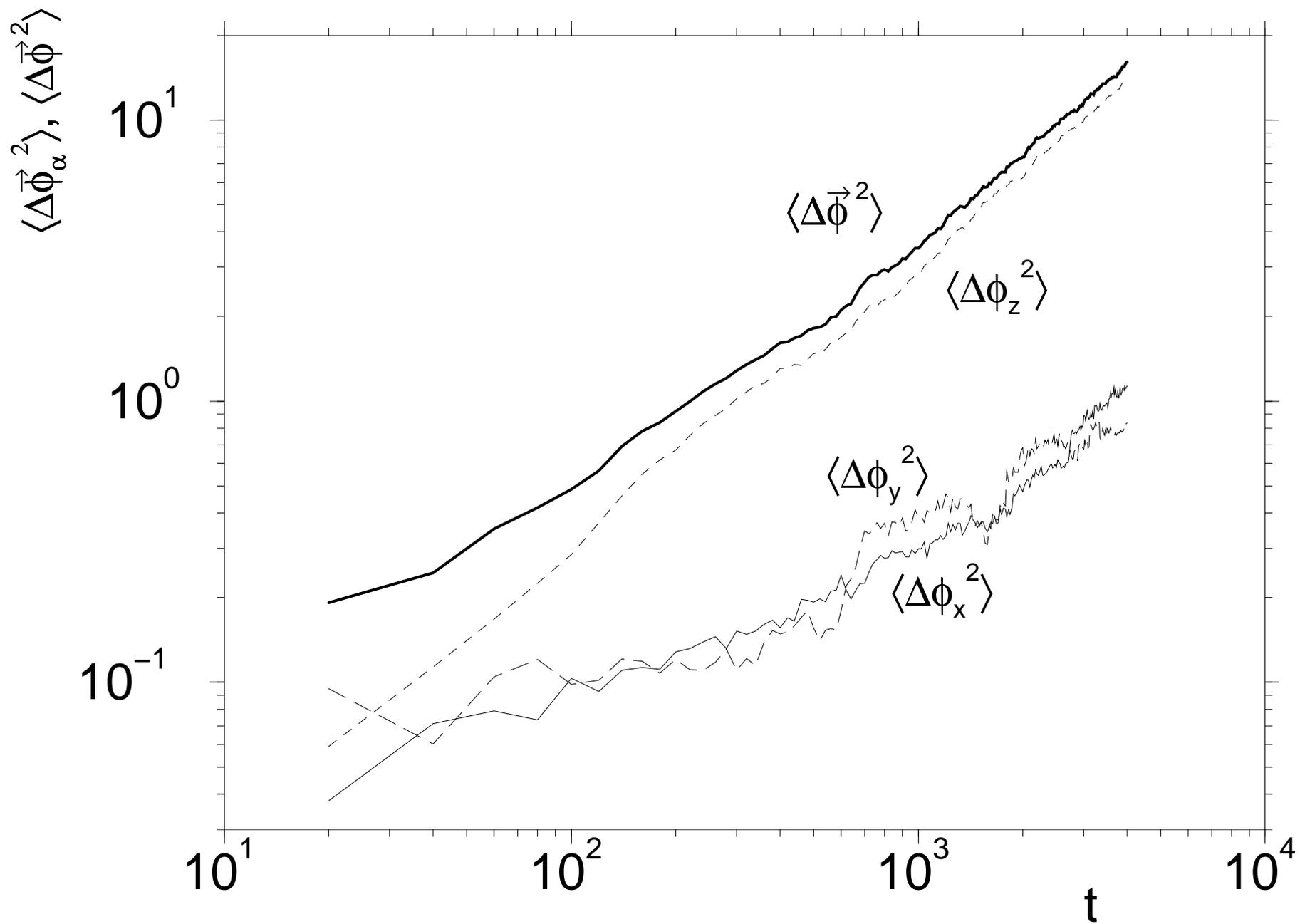
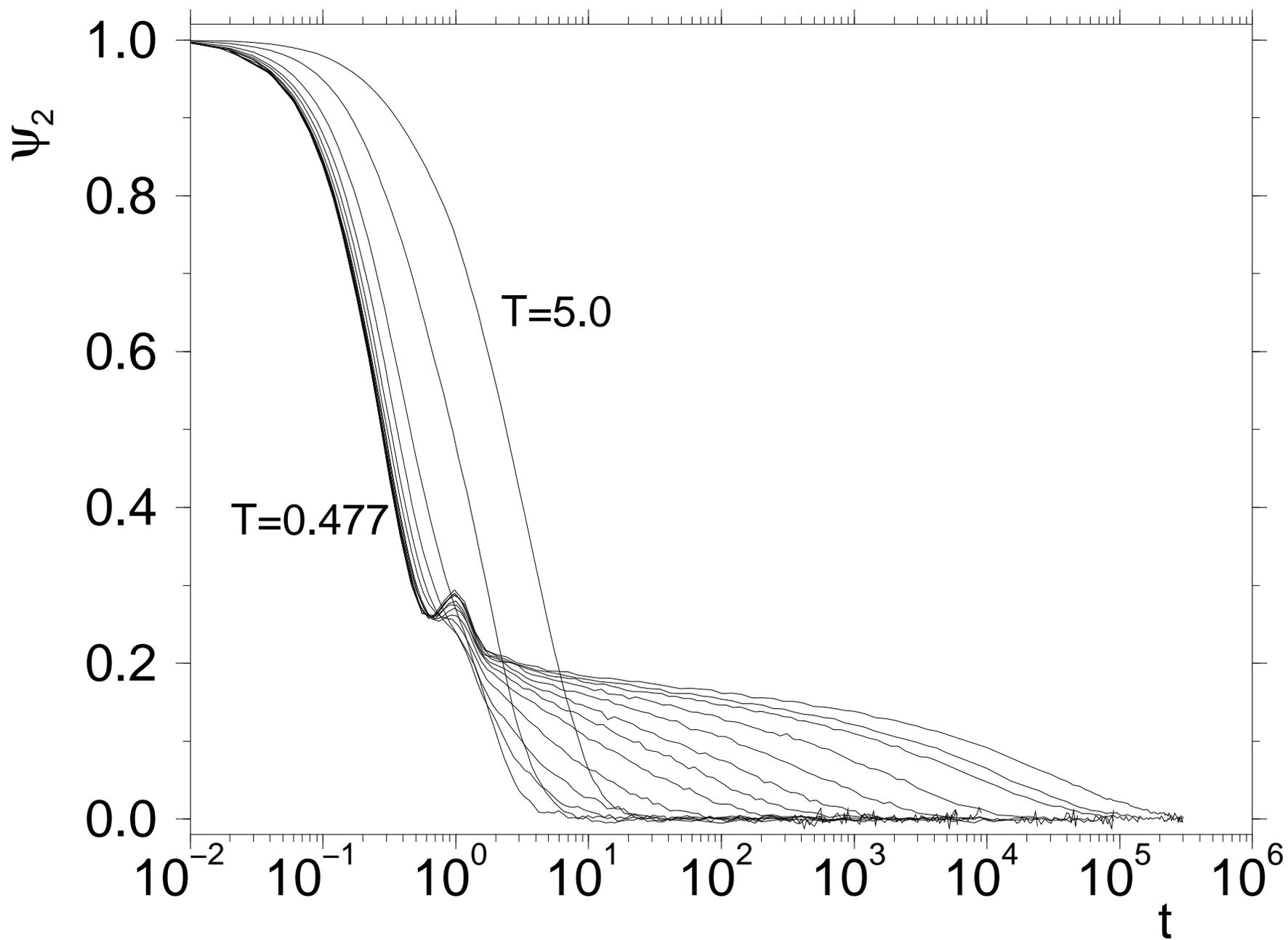
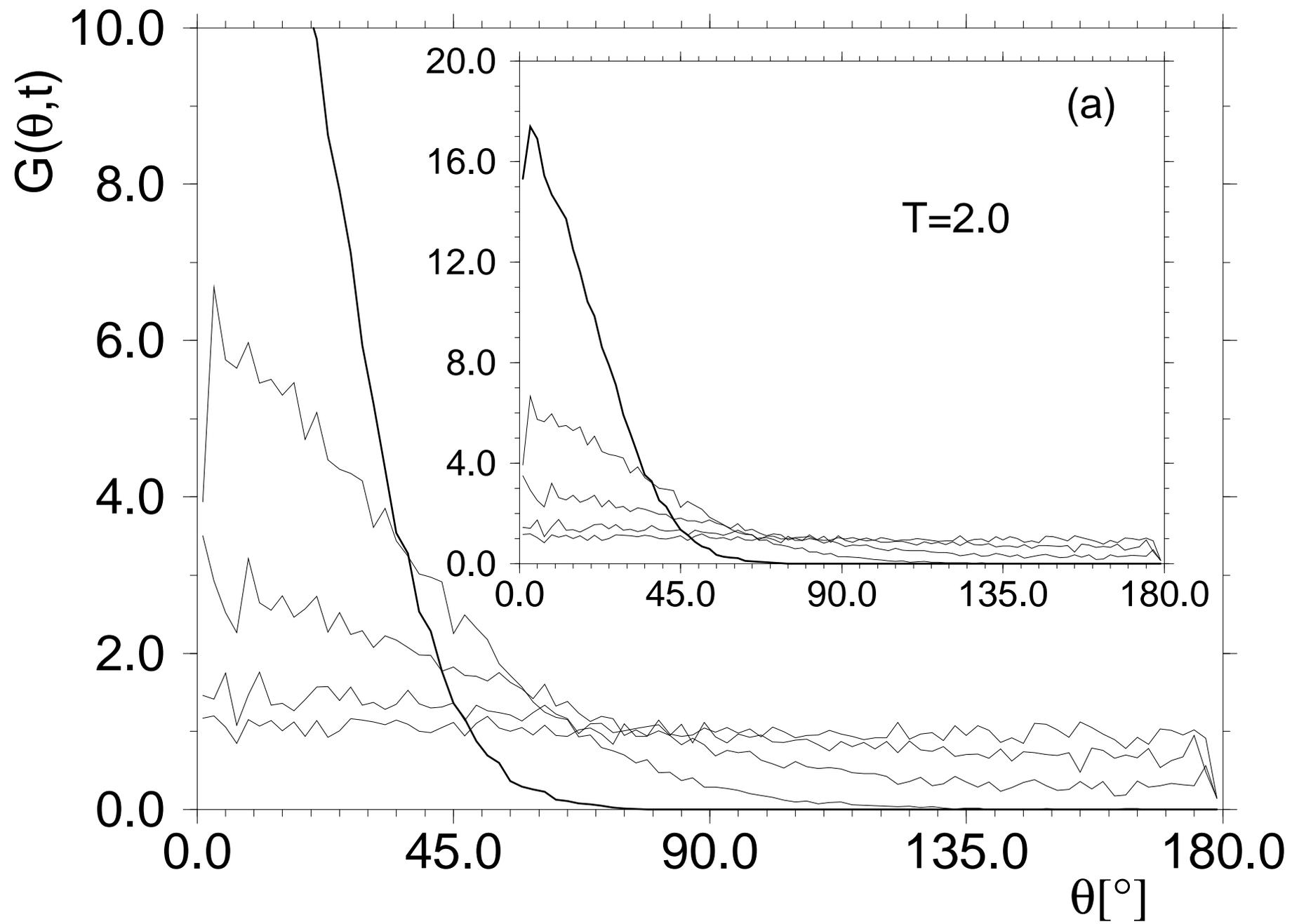
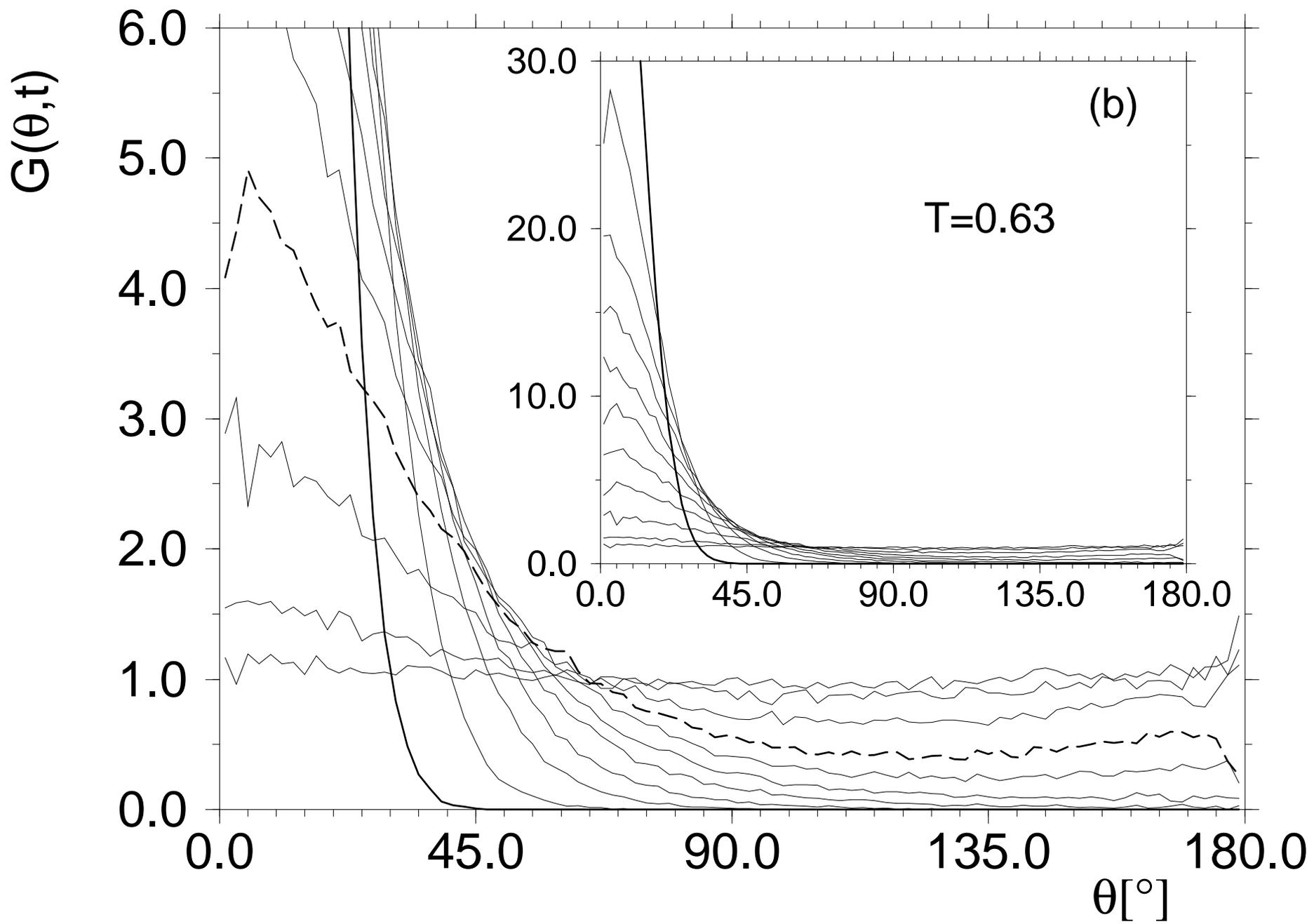
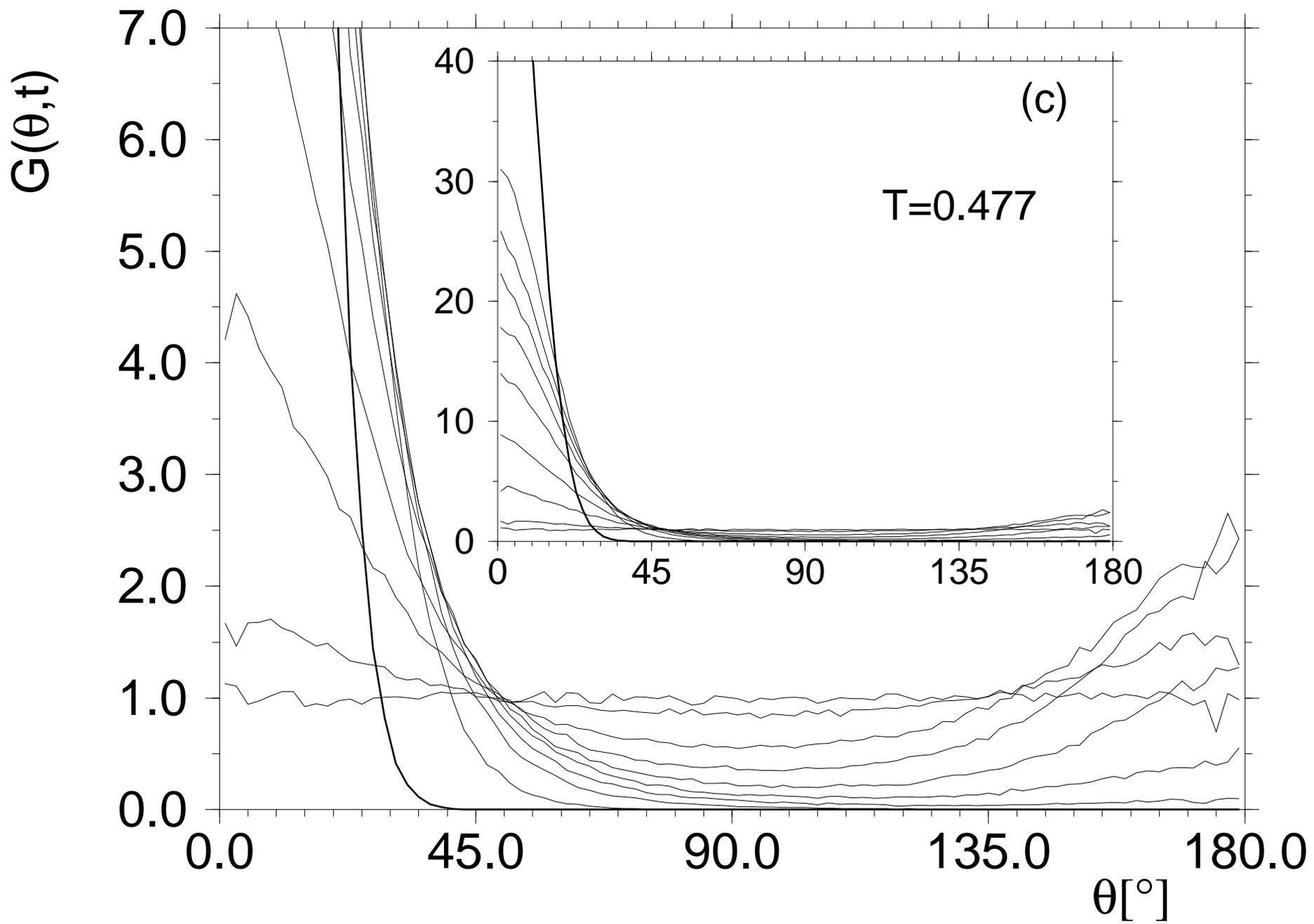
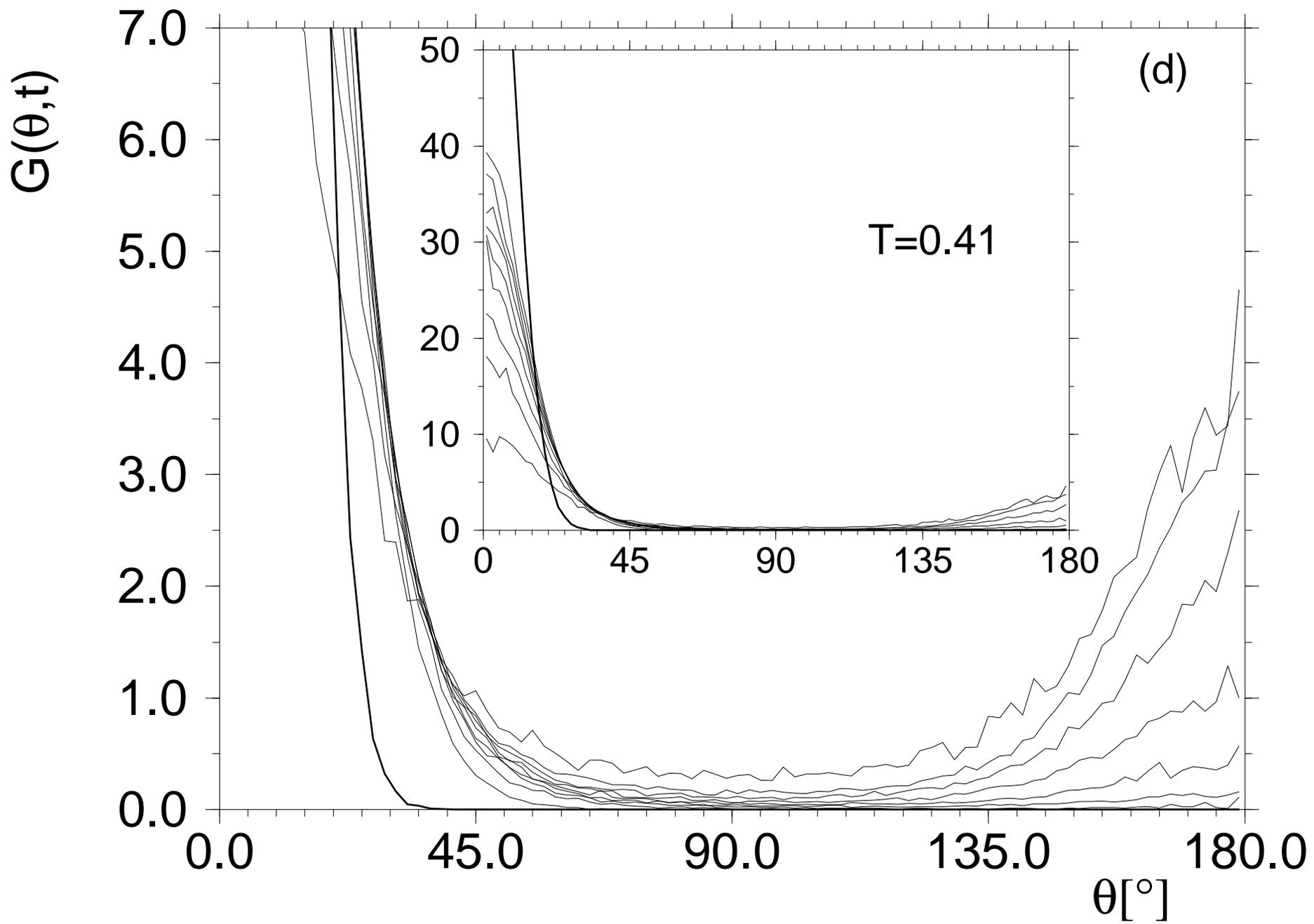
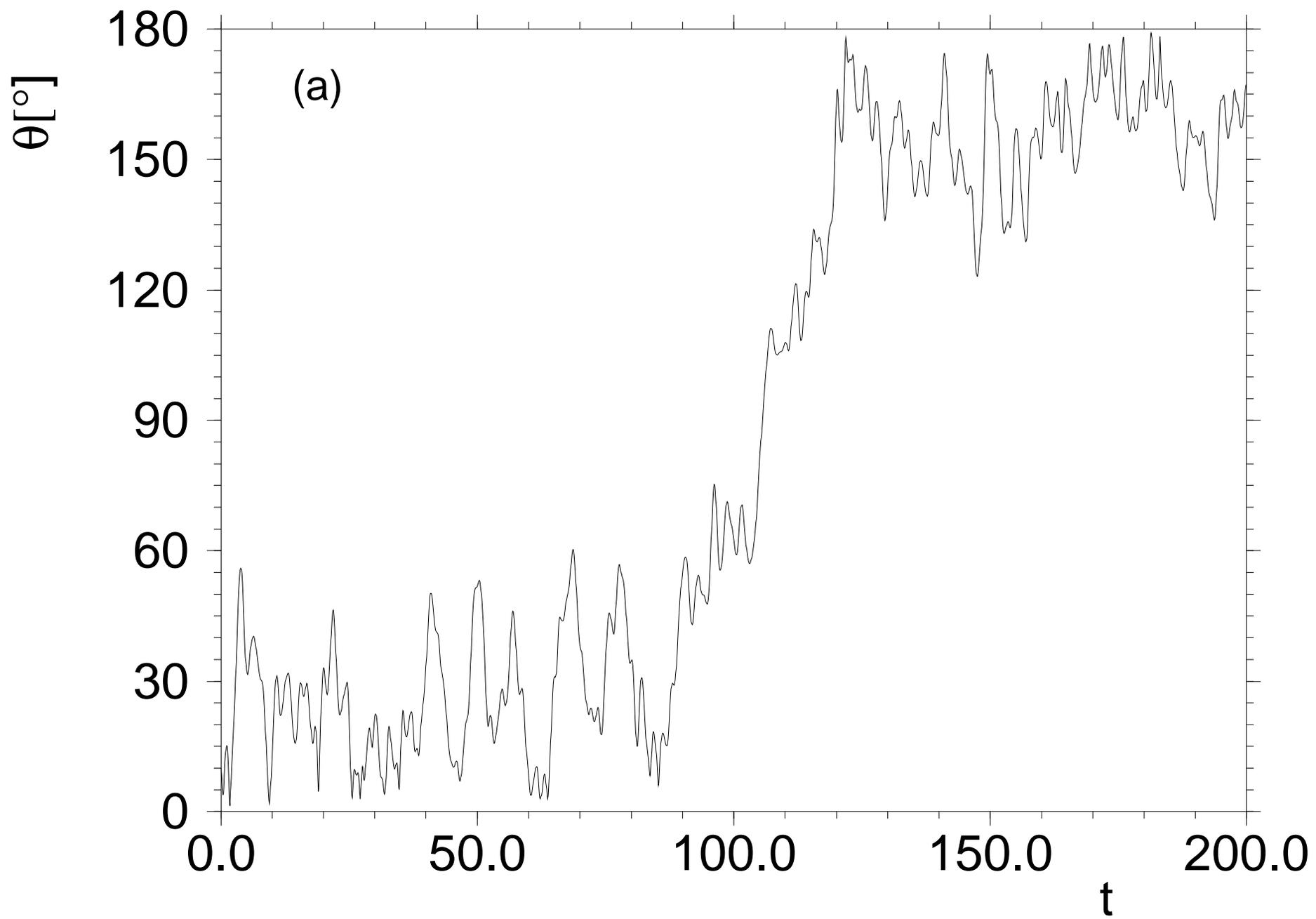
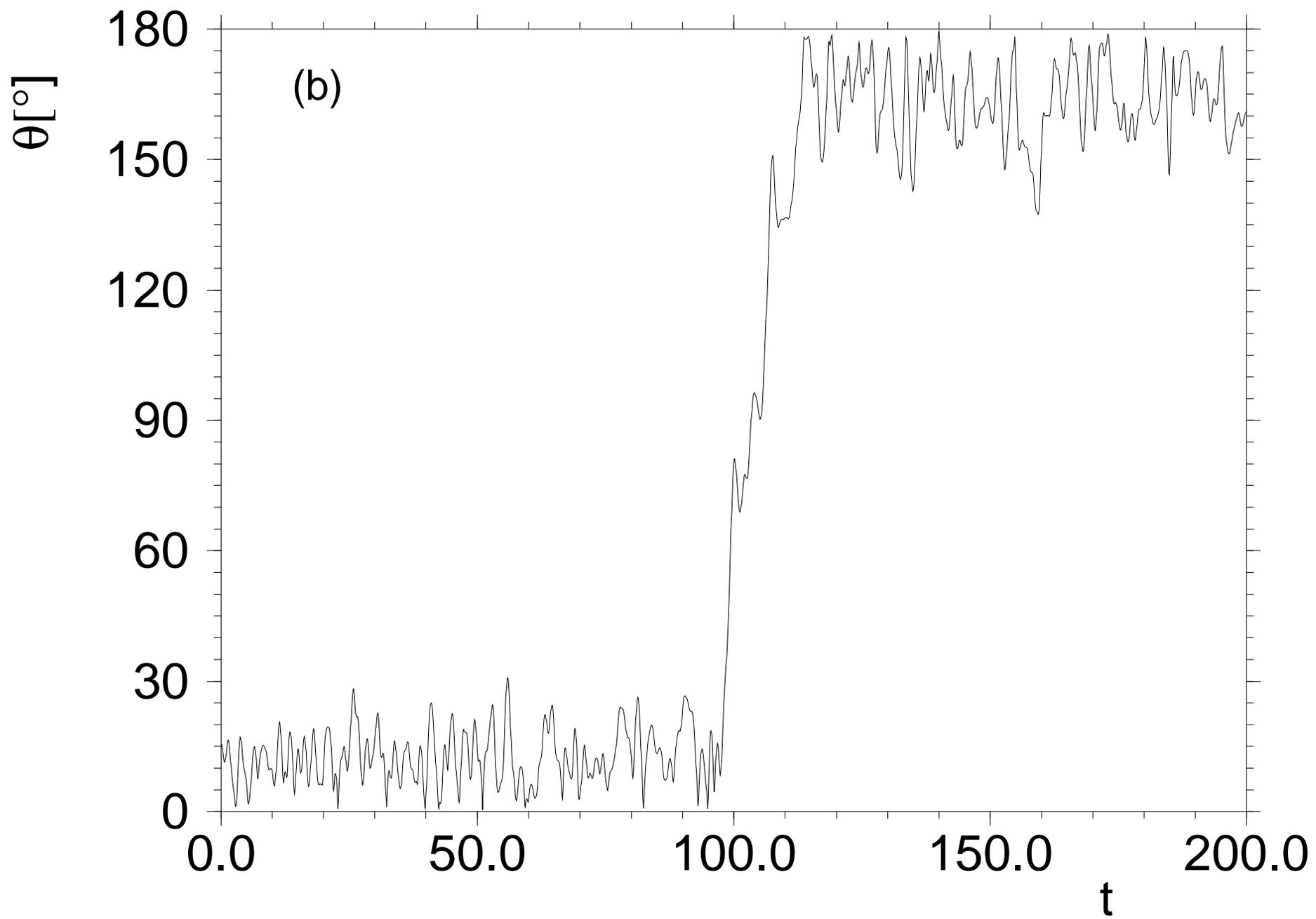
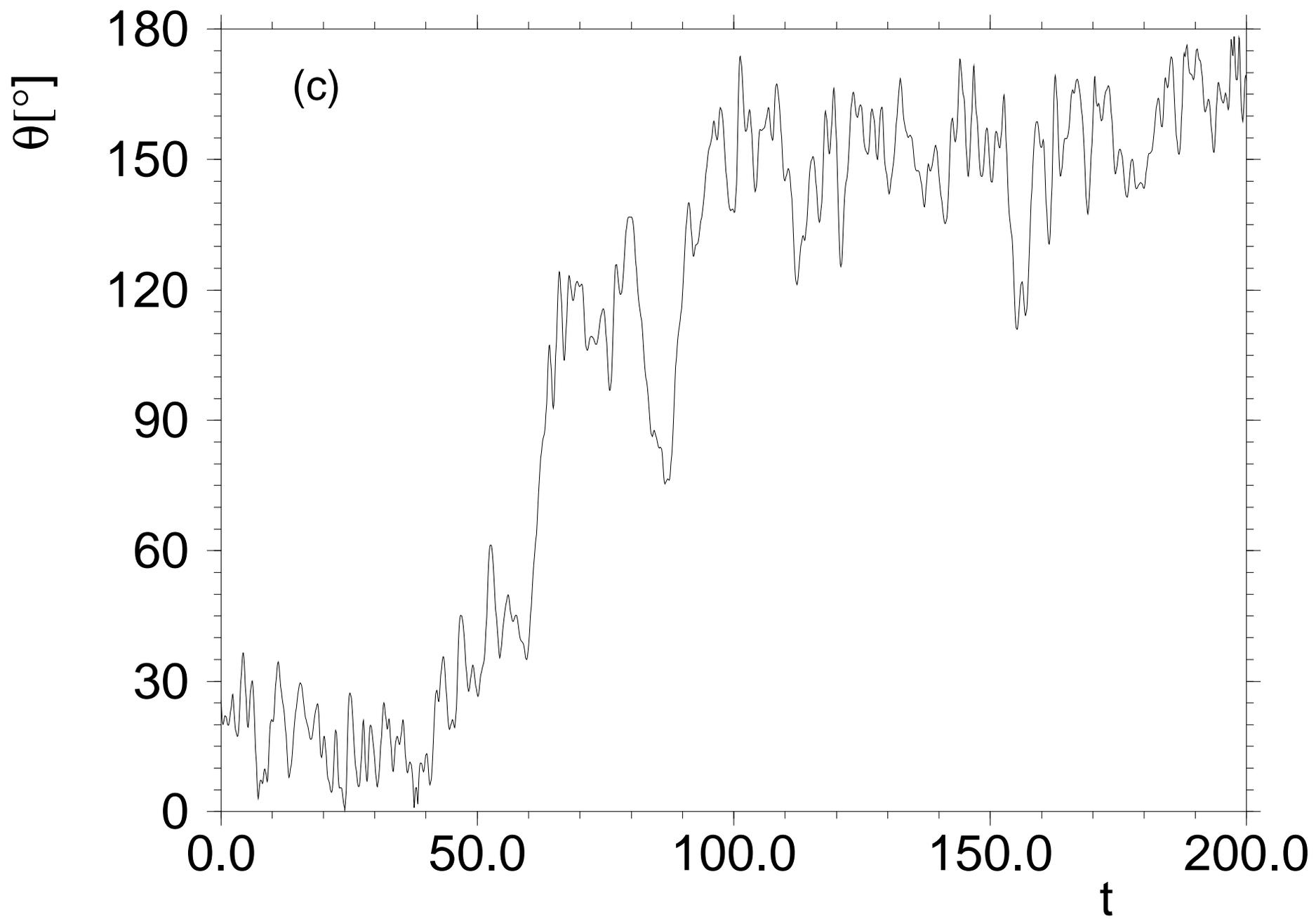
\end{document}